\documentclass[runningheads]{llncs}

\usepackage[table,dvipsnames]{xcolor}
\usepackage{hyperref}
\usepackage{caption}
\usepackage{amsmath}
\usepackage{amssymb}
\usepackage[capitalize]{cleveref}
\usepackage{xspace}
\usepackage{tikz}
\usepackage{subcaption}
\usepackage{wrapfig}
\usepackage{booktabs}
\usepackage{siunitx} % for formatting numbers in tables

\usepackage{newtxmath}
\usepackage{newtxtext}

\usetikzlibrary{shapes.misc, positioning}
\usetikzlibrary{arrows,automata}

\newcommand{\ol}[1]{\textcolor{blue}{\ifmmode \text{[OL: #1]}\else [OL: #1] \fi}}
\newcommand{\lh}[1]{\textcolor{red!70}{\ifmmode \text{[LH: #1]}\else [LH: #1] \fi}}
\newcommand{\js}[1]{\textcolor{purple}{\ifmmode \text{[JS: #1]}\else [JS: #1] \fi}}
\newcommand{\vh}[1]{\textcolor{green!70!black}{\ifmmode \text{[VH: #1]}\else [VH: #1] \fi}}
\newcommand{\yfc}[1]{\textcolor{blue!70!black}{\ifmmode \text{[YFC: #1]}\else [YFC: #1] \fi}}

\newcommand{\varss}{\mathbb{X}_s}
\newcommand{\varsi}{\mathbb{X}_i}

\makeatletter
\DeclareFontEncoding{LS1}{}{}
\makeatother
\DeclareFontSubstitution{LS1}{stix}{m}{n}
\DeclareSymbolFont{stixletters}{LS1}{stix}{m}{it}
\DeclareSymbolFont{stixoperators}{LS1}{stix}{m}{n}

\DeclareMathSymbol{\othersubseteq}{\mathrel}{stixletters}{"D3}
\DeclareMathSymbol{\otherin}{\mathrel}{stixoperators}{"CB}
\DeclareMathSymbol{\othereq}{\mathrel}{stixoperators}{`=}

\renewcommand{\prod}{\mathit{Product}}

\newcommand{\toolfont}[1]{\textsc{#1}}
\newcommand{\mata}{\toolfont{Mata}\xspace}
\newcommand{\ziiinoodler}{\toolfont{Z3-Noodler}\xspace}
\newcommand{\ziiinoodleroopsla}{\toolfont{Z3-Noodler}$^{\mathit{pr}}$\xspace}
\newcommand{\noodler}{\toolfont{Noodler}\xspace}
\newcommand{\cvc}{\toolfont{cvc}\xspace}
\newcommand{\cvcv}{\cvc{}5\xspace}

\newcommand{\ziii}{\toolfont{Z3}\xspace}
\newcommand{\ziiistriiire}{\toolfont{Z3str3RE}\xspace}
\newcommand{\ziiistriv}{\toolfont{Z3str4}\xspace}
\newcommand{\ziiitrau}{\toolfont{Z3-Trau}\xspace}

\newcommand{\ostrich}[0]{\toolfont{OSTRICH}\xspace}

\newcommand{\vbsnoodler}[0]{\toolfont{VBS}$^+$\xspace}

\newcommand{\sumcol}[1]{\textit{#1}}
\newcommand{\benchfont}[1]{\textsf{#1}}
\newcommand{\nornbench}{\benchfont{Norn}\xspace}
\newcommand{\slog}{\benchfont{Slog}\xspace}
\newcommand{\slent}{\benchfont{Slent}\xspace}
\newcommand{\sygusqgen}{\benchfont{Sygus-qgen}\xspace}
\newcommand{\leetcode}{\benchfont{LeetCode}\xspace}
\newcommand{\kaluza}{\benchfont{Kaluza}\xspace}
\newcommand{\woorpje}{\benchfont{Woorpje}\xspace}
\newcommand{\keplerbench}{\benchfont{Kepler}\xspace}
\newcommand{\fullstrint}{\benchfont{FullStrInt}\xspace}
\newcommand{\pyex}{\benchfont{PyEx}\xspace}
\newcommand{\strsmall}{\benchfont{StrSmallRw}\xspace}
\newcommand{\stringfuzz}{\benchfont{StringFuzz}\xspace}
\newcommand{\webapp}{\benchfont{Webapp}\xspace}
\newcommand{\denghang}{\benchfont{Denghang}\xspace}
\newcommand{\transducerplus}{\benchfont{Transducer+}\xspace}
\newcommand{\automatark}{\benchfont{AutomatArk}\xspace}

\newcommand{\nornbenchshort}{\benchfont{Norn}\xspace}
\newcommand{\slogshort}{\benchfont{Slog}\xspace}
\newcommand{\slentshort}{\benchfont{Slent}\xspace}
\newcommand{\sygusqgenshort}{\benchfont{Syg}\xspace}
\newcommand{\leetcodeshort}{\benchfont{Leet}\xspace}
\newcommand{\kaluzashort}{\benchfont{Kal}\xspace}
\newcommand{\woorpjeshort}{\benchfont{Woo}\xspace}
\newcommand{\keplerbenchshort}{\benchfont{Kep}\xspace}
\newcommand{\fullstrintshort}{\benchfont{StrInt}\xspace}
\newcommand{\pyexshort}{\benchfont{PyEx}\xspace}
\newcommand{\strsmallshort}{\benchfont{StrSm}\xspace}
\newcommand{\stringfuzzshort}{\benchfont{StrFuzz}\xspace}
\newcommand{\webappshort}{\benchfont{Web}\xspace}
\newcommand{\denghangshort}{\benchfont{Den}\xspace}

\newcommand{\automatarkshort}{\benchfont{Aut}\xspace}

\newcommand{\benchcategoryfont}[1]{\textbf{#1}}
\newcommand{\regexbench}{\benchcategoryfont{Regex}\xspace}
\newcommand{\eqbench}{\benchcategoryfont{Equations}\xspace}
\newcommand{\predbench}{\benchcategoryfont{Predicates-small}\xspace}

\newcommand{\QFS}{\texttt{QF\_S}\xspace}
\newcommand{\QFSLIA}{\texttt{QF\_SLIA}\xspace}

\newcommand{\RGBcircle}[1]{\ensuremath{{\color[RGB]{#1}\bullet}}}

\def\markbaseline{-.28em}   % from spot
\newcommand{\colcircle}[1]{node[left,shape=circle,inner sep=1pt,scale=0.7,fill=orange!60!black,text=white,draw=orange!60!black]  {\sf #1}}
\newcommand{\textcolcircle}[1]{\tikz[baseline=\markbaseline,scale=0.5]{ \draw \colcircle{#1};}}

\newcommand{\regex}{\mathcal{R}}

\newcommand{\smtlib}[0]{SMT-LIB\xspace}

\newcommand{\strpr}[1]{\texttt{#1}\xspace}
\newcommand{\substr}{\strpr{substr}}
\newcommand{\at}{\strpr{at}}
\newcommand{\replace}{\strpr{replace}}
\newcommand{\replaceall}{\strpr{replace\_all}}
\newcommand{\replacere}{\strpr{replace\_re}}
\newcommand{\replacereall}{\strpr{replace\_re\_all}}
\newcommand{\indexof}{\strpr{indexof}}
\newcommand{\contains}{\strpr{contains}}
\newcommand{\notcontains}{\neg\contains}
\newcommand{\prefix}{\strpr{prefix}}
\newcommand{\suffix}{\strpr{suffix}}

\newcommand{\toint}{\strpr{to\_int}}
\newcommand{\fromint}{\strpr{from\_int}}

\newcommand{\len}[0]{\strpr{len}}
\newcommand{\lenof}[1]{\len(#1)}

\InputIfFileExists{cutmargins.tex}{%
	\pdfpagesattr{/CropBox [125 82 490 688]} % LNCS page made even biger
}{%
}

\title{\ziiinoodler: An Automata-based String Solver
\\ (Technical Report)
}
\author{
Yu-Fang Chen\inst{1} \and
David Chocholatý\inst{2} \and
Vojtěch Havlena\inst{2} \and \\
Lukáš Holík\inst{2} \and
Ondřej Lengál\inst{2} \and
Juraj Síč\inst{2}
}

\institute{
Academia Sinica, Taipei, Taiwan \and
Brno University of Technology, Brno, Czech Republic 
}

\begin{document}
\maketitle

% remove to suppress page numbering
\pagestyle{plain}

\begin{abstract}
  \vspace{-3mm}
  \ziiinoodler is a~fork of \ziii that replaces its string theory solver with
  a~custom solver implementing the recently introduced stabilization-based
  algorithm for solving word equations with regular constraints.
  An extensive experimental evaluation shows that
  \ziiinoodler is a~fully-fledged solver that can compete with
  state-of-the-art solvers, surpassing them by far on many benchmarks.
  Moreover, it is often complementary to other solvers, making it a~suitable
  choice as a~candidate to a~solver portfolio.
  % For this, we needed to develop and implement heuristics driven by real-world
  % benchmarks.
  % \ol{todo}
  %
  % \ziiinoodler is a~fork of \ziii that replaces its string theory solver with
  % a~custom solver implementing the recently introduced stabilization-based
  % algorithm for solving word equations with regular constraints.
  % An extensive experimental evaluation shows that
  % \ziiinoodler has reached a~quite mature state, being able to compete with
  % state-of-the-art solvers, in many cases surpassing them, 
\end{abstract}

%%%%%%%%%%%%%%%%%%%%%%%%%%%%%%%%%%%%%%%%%%%%%%%%%%%%%%%%%%%%%%%%%%
\vspace{-8mm}
\section{Introduction}
\vspace{-2mm}
%%%%%%%%%%%%%%%%%%%%%%%%%%%%%%%%%%%%%%%%%%%%%%%%%%%%%%%%%%%%%%%%%%
Recently, many tools for solving string constraints have been developed,
motivated mainly by techniques for finding security vulnerabilities such as SQL
injection or cross-site scripting (XSS) in web
applications~\cite{OWASP13,OWASP17,OWASP21}.
String solving has also found its applications in, e.g., analysis of
access user policies in Amazon Web
Services~\cite{hadarean_mosca,stringsAWS18,Rungta2022} or smart
contracts~\cite{AltBHS22}.
% verification of programs manipulating strings, especially in the context of web programs where they are used to detect vulnerabilities such as SQL injection or cross-site scripting (XSS) attacks, which are still considered to be among the most frequent and problematic sources of security bugs related to web applications~\cite{OWASP13,OWASP17,OWASP21}.
Solvers for string constraints are usually implemented as string theory solvers
inside SMT solvers, such as \cvcv~\cite{cvc5} or \ziii~\cite{z3}, allowing
combination with other theories, most commonly the theory of integers for
string lengths.
Amonong well known string solvers other than than \cvcv and \ziii are
\ziiistriiire~\cite{Z3str3RE,BerzishDGKMMN23},
\ziiitrau~\cite{abdulla_efficient_2020}, \ziiistriv~\cite{MoraBKNG21},
\ostrich~\cite{AnthonyComplex2019}, and there are many more.

In this paper, we present \ziiinoodler~1.0.0~\cite{Z3Noodler}, a~fork of
\ziii~4.12.2 where the string theory
solver is replaced with the \emph{sta\-bi\-liza\-tion-based procedure} for
solving string (dis)equations with regular and length
constraints~\cite{NoodlerFM23,NoodlerOOPSLA23}.
The procedure makes heavy use of \emph{nondeterministic finite
automata} (NFAs) and operations over them, for which we use the efficient \mata
library for NFAs~\cite{MataPaper,Mata}.

% In this paper we present version \js{1.0.0?} of another string solver called \ziiinoodler\footnote{The tool is available at \url{https://github.com/VeriFIT/z3-noodler}.}, first introduced in~\cite{NoodlerOOPSLA23}.
% It is based on SMT solver \ziii whose string theory solver is replaced with the \emph{sta\-bi\-liza\-tion-based procedure} for solving string equations and regular constraints~\cite{NoodlerFM23} extended with handling of length constraints and disequations~\cite{NoodlerOOPSLA23}.
% It uses finite automata to represent regular constraints and the decision procedure works with these automata.
% We use library \mata (previously known as \toolfont{eNFA}~\cite{fiedor2023reasoning,NoodlerOOPSLA23}) \js{cite arxiv when we put mata paper there} for implementation of automata and operations over them.

The presented version
implements multiple improvements over a~previous
\ziiinoodler prototype from~\cite{NoodlerOOPSLA23}.
Firstly, it extends the support for string predicates from the \smtlib string
theory standard~\cite{SMTLIB-Strings} by
(1)~applying smarter and more specific axiom saturation and
(2)~adding support for their solving inside the decision procedure
% (\texttt{to}/\fromcode and $\notcontains$ predicates).
(e.g., for the $\notcontains$ predicate).
It also implements various optimizations (e.g., for regular constraints handling) and other decision procedures, e.g., the \emph{Nielsen
transformation}~\cite{nielsen1917} for quadratic equations or a~procedure for
regular language (dis)equations;
moreover, we added heuristics for choosing the best decision procedure to use.
% \ol{fixed?}
% \lh{this sounds like it was not selected automatically before, which is fishy}
% (including an incomplete decision procedure from~\cite{NoodlerOOPSLA23} where suitable).
% Furthermore, it can now automatically select which decision procedure should be used based on the formula.
% Other than the newly implemented decision procedures, this can also turn on the
% underapproximating decision procedure from~\cite{NoodlerOOPSLA23} when it
% detects that it would be better for solving.
% Previously, underapproximation had to be turned on manually.
%
% Finally, it also benefits from many improvements in \mata such as faster implementation of automata operations.

We compared \ziiinoodler with other string solvers on standard \smtlib
benchmarks~\cite{SMTLIB,QF_S,QF_SLIA}.
The results indicate that \ziiinoodler is competitive, superior especially on
benchmarks containing mostly regular constraints and word (dis)equations, and
that the improvements since~\cite{NoodlerOOPSLA23} had a~large impact on the
number of solved instances as well as its overall performance.
%, even fixing multiple bugs.

% Furthermore, they also show that the improvements since~\cite{NoodlerOOPSLA23}
% had large impact on the number of solved instances, even fixing multiple bugs.

% To show the strengths of \ziiinoodler, we compared it with other string solvers on standard \smtlib benchmarks~\cite{SMTLIB,QF_S,QF_SLIA}.
% The results indicate that \ziiinoodler is indeed very competitive solver, especially on benchmarks containing mostly regular constraints and string (dis)equations.
% Furthermore, they also show that the improvements since~\cite{NoodlerOOPSLA23} had large impact on the number of solved instances, even fixing multiple bugs.

%%%%%%%%%%%%%%%%%%%%%%%%%%%%%%%%%%%%%%%%%%%%%%%%%%%%%%%%%%%%%%%%%%
\vspace{-2mm}
\section{Architecture}\label{sec:architecture}
\vspace{-1mm}
%%%%%%%%%%%%%%%%%%%%%%%%%%%%%%%%%%%%%%%%%%%%%%%%%%%%%%%%%%%%%%%%%%

\ziiinoodler replaces the string theory solver in the
DPLL(T)-based SMT solver \ziii~\cite{z3} (version~4.12.2) by our solver
\noodler~\cite{NoodlerFM23}, which is based on the \emph{stabilization algorithm} (cf.~\cref{sec:dec-proc}).
DPLL(T)-based solvers in general combine a~SAT solver providing satisfying assignments 
to the Boolean skeleton of a~formula with multiple theory solvers for checking
conjunctions of theory literals.
% \ziiinoodler implements the \emph{lazy approach}, meaning that it checks satisfiability of
% a~conjunction of string atoms after the SAT solver constructs a~satisfying assignment of the 
% Boolean skeleton.
%

%Although \ziiinoodler replaces the string theory solver in \ziii, 
\ziiinoodler still uses the infrastructure 
of \ziii, most importantly the parser, string theory rewriter and the \emph{linear integer arithmetic} (LIA) solver.
The \ziii parser takes formulae in the SMT-LIB format~\cite{SMTLIB}, where
\ziiinoodler can %therefore parse formulae in SMT format and string theory.
handle nearly all predicates/functions (such as \substr, \len, \at, \replace, regular membership, word equations, etc.) in the string theory as defined by SMT-LIB~\cite{SMTLIB-Strings}.
% It cannot handle string-int conversions, lexicographic comparison, regular expressions containing variables, \replaceall, and replacements based on regular expressions.
See \cref{sec:string-logic} for a~definition of the fragment that we support.
% \js{At the end of \cref{sec:string-theory}, there is a paragraph about what we can handle, maybe the previous two sentences should not be here then}.

Even though we do use the string theory rewriter of \ziii, we disabled those rewritings that do not benefit our core string solver.  
%, some of the rewritings are undesirable and are disabled in \ziiinoodler.
For instance, we removed rules that rewrite regular membership constraints to other types of constraints since solving regular constraints and word equations using our stabilization-based approach is efficient.

\begin{wrapfigure}[10]{r}{5cm}
  \captionsetup{font=small}
    \vspace*{-9mm}
    \hspace*{-4mm}
    \begin{minipage}{5.6cm}
    \scalebox{0.73}{
    \begin{tikzpicture}[
  auto,
  transform shape,
  node distance=20mm
]

% z3 box
\draw [draw]
  (0,0) --
  ++(2.5,0) --
  ++(0,-1.25) --
  ++(2.5,0) {[rounded corners=7] --
  ++(0,3) --
  ++(-5,0)} --
  cycle
  {};

% noodler box
\draw [draw, fill=orange!20]
  (0,-0.1) --
  ++(2.4,0) --
  ++(0,-2.4) {[rounded corners=7] --
  ++(-2.4,0)} --
  cycle{} node[align=center,text width=2cm] at (1.2,-1.3) {\noodler \textsf{string theory}};

% lia external box
\draw [draw]
  (2.5, -1.35)  --
  ++(2.5,0) {[rounded corners=7] --
  ++(0,-1.15)}  --
  ++(-2.5,0) --
  cycle {} node[align=center,text width=2cm] at (3.8,-2) { \textsf{LIA solver instance}};

% mata box
\draw [draw]
  (-1.2, 1.75) {[rounded corners=7] --
  ++(1,0) --
  ++(0,-4.25)  --
  ++(-1,0) --
  cycle} {} node[align=center,text width=2cm] at (-0.72,-0.5) { \textsc{Mata}};

% lia solver core
\draw [draw, gray, dashed]
  (2.6, 0.0) {[rounded corners=5]  --
  ++(2.3,0)  --
  ++(0,-1.15)  --
  ++(-2.3,0) --
  cycle } {} node[align=center,text width=2cm] at (3.8,-0.5) { \textsf{LIA solver}};

% z3 core
\draw [draw, gray, dashed]
  (0.1, 1.65) {[rounded corners=5]  --
  ++(4.8,0)  --
  ++(0,-1.55)  --
  ++(-4.8,0) --
  cycle } {} node[align=center,text width=2cm] at (1.4,0.9) { \textsf{core}};

% theory rewriter
\draw [draw, gray, dashed]
  (2.6, 1.55) {[rounded corners=5]  --
  ++(2.2,0)  --
  ++(0,-1.35)  --
  ++(-2.2,0) --
  cycle } {} node[align=center,text width=1.5cm] at (3.8,0.9) { \textsf{string rewriter}} ;

% arrows
\draw[->,stealth-] (0.8,-0.3) -- (0.8,0.3);
\draw[->,-stealth] (1.8,-0.3) -- (1.8,0.3);
\draw[->,stealth-stealth] (2.15,-2.1) -- (2.75,-2.1);
\draw[->,stealth-stealth] (-0.4,-2.1) -- (0.2,-2.1);
\draw[->,-stealth] (2.5,2) -- (2.5,1.4);

% smt formula
\node (smt-formula) at (2.5, 2.3) {\textsf{SMT string formula}};

\node [scale=2.0] (preds) at (0.5,0.0) {\textcolcircle{1}};
\node [scale=2.0] (lemmas) at (1.5,0.0) {\textcolcircle{2}};
\node [scale=2.0] (lengths) at (2.45,-1.8) {\textcolcircle{4}};
\node [scale=2.0] (automata) at (-0.1,-1.8) {\textcolcircle{3}};

\node (ziii) at (0.5, 2.0) {\textsf{Z3}};

\end{tikzpicture}
    }
    \end{minipage}
    \vspace{-3mm}
    \caption{Architecture of \ziiinoodler}
   \label{fig:architecture}
\end{wrapfigure}

The interaction of the \noodler solver with \ziii is shown in
\cref{fig:architecture} and works as follows.
Upon receiving a~satisfying Boolean assignment from the SAT
solver~(\textcolcircle{1}), we first remove irrelevant assignments 
(using \ziii's relevancy propagation), which allows us
to work with smaller instances and return more general theory lemmas.
An assignment consists of string (dis)equations, regular constraints, and,
possibly, predicates that were not \emph{axiom-saturated} before
(cf.~\cref{sec:axiomatization}).
%\js{should we explain what axiomatize means?}.
%We axiomatize predicates such as \texttt{indexof} or \texttt{substr} using string (dis)equations with length and regular constraints
%However, this is not always posssible, 
%for instance for general $\neg\mathtt{contains}$ predicates. These predicates we handle separately inside the decision procedures.

The core \noodler string decision procedure then reduces the conjunction of string literals to a LIA constraint over string lengths, 
and returns it to \ziii as a~theory lemma~(\textcolcircle{2}), to be solved together with the rest of the input arithmetic constraints by \ziii{'s} internal LIA solver.
\noodler implements a couple of decision procedures (discussed in \cref{sec:string-theory}), heavily employing the \mata automata library (version 0.109.0)~\cite{Mata}~(\textcolcircle{3}).
As an optimization of the theory lemma generation, 
when the string constraint reduces into a~disjunction of LIA length constraints, 
we check the satisfiability of individual disjuncts (generated lazily on
demand) separately in order to get a positive answer as soon as possible. 
For testing the disjuncts, the current solver context is cloned and queried
about satisfiability of the LIA constraint conjoined with the disjunct~(\textcolcircle{4}).
% The disjuncts are generated lazily on demand.

%%%%%%%%%%%%%%%%%%%%%%%%%%%%%%%%%%%%%%%%%%%%%%%%%%%%%%%%%%%%%%%%%%
\vspace{-2.0mm}
\section{String Theory Core}
\vspace{-1.0mm}
\label{sec:string-theory}
%%%%%%%%%%%%%%%%%%%%%%%%%%%%%%%%%%%%%%%%%%%%%%%%%%%%%%%%%%%%%%%%%%
In this section, we provide details about \ziiinoodler's string 
theory implementation.

%%%%%%%%%%%%%%%%%%%%%%%%%%%%%%%%%%%%%%%%%%%%%%%%%%%%%%%%%%%%%%%%%%
% \subsection{Axiom Saturation}\label{sec:axiomatization}
%%%%%%%%%%%%%%%%%%%%%%%%%%%%%%%%%%%%%%%%%%%%%%%%%%%%%%%%%%%%%%%%%%

\vspace{-1mm}
\paragraph{Axiom Saturation.}\label{sec:axiomatization}
In order to best utilize the power of \ziii's internal LIA solver during the generation
of a~satisfiable assignment, we saturate the input formula with length-aware
theory axioms and axioms for string predicates (this happens during \ziii's
processing of the input formula, before the main SAT solver starts generating
assignments).
We can then avoid checking SAT assignments that 
trivially violate length conditions.
Most importantly, we add length axioms $\lenof{t_1} \geq 0$, $\lenof{t_1.t_2} =
\lenof{t_1} + \lenof{t_2}$ where $t_1, t_2$ are arbitrary string terms, and
$\lenof{t_1} = \lenof{t_2}$ for the word equation $t_1 = t_2$.
% In a~similar fashion, we axiomatize string predicates breaking it to a~series
% of string (dis)equations, length, and regular constraints.
% This is, however, not always possible (e.g., for $\notcontains$ or the extended constraint $\toint$).
% Therefore, we handle these predicates separately inside the decision procedure itself\footnote{In the current version of \ziiinoodler, we support only a~limited form of $\notcontains$ from non-axiomatized predicates.}. 
% \js{The following was moved to \cref{sec:architecture} but it was not correct, rewriting actually happens during the call of \ziii to string sovler where it informs it that some string predicate/function exists.}

% \vspace{-2mm}
%------------------------------------------------------------------------------
% \paragraph{Custom Rewriting.}

Moreover, for string functions/predicates, \noodler saturates the original
formula with an equivalent formula composed 
% As the first step taken after receiving an assignment,
% \noodler rewrites some string predicates and function/predicate calls into an equivalent simpler 
% series
of word (dis)equations and length/regular constraints, which are more suitable
for our core procedure (e.g., for $\neg\contains(s, \strpr{"abc"})$ in the
input formula, we add the regular constraint $s \notin \Sigma^* \strpr{abc}
\Sigma^*$).
We use different saturation rules for instances of predicates with concrete
values.
For instance, for $\substr(s,\strpr{4},\strpr{1})$, we add just the formula $\at(s, \strpr{4})$.
On the other hand, for $\substr(s,t_i,t_j)$, where~$s$ is a string term and $t_i, t_j$ are
general integer terms (possibly containing variables), we need to add a~more
general formula talking about the prefix and suffix of~$s$ of given lengths.
The original predicate occurrence is then removed from received assignments by \noodler (\ziii does not allow to remove parts of the original formula).

%%%%%%%%%%%%%%%%%%%%%%%%%%%%%%%%%%%%%%%%%%%%%%%%%%%%%%%%%%%%%%%%%%
% \subsection{Decision Procedures}\label{sec:dec-proc}
%%%%%%%%%%%%%%%%%%%%%%%%%%%%%%%%%%%%%%%%%%%%%%%%%%%%%%%%%%%%%%%%%%

%!!!!!!!!!!!!!!!!!!!!!!!!!!!!!!!!
\enlargethispage{1mm}
%!!!!!!!!!!!!!!!!!!!!!!!!!!!!!!!!

\vspace{-3mm}
\paragraph{Decision Procedures.}\label{sec:dec-proc}

\ziiinoodler's string theory core contains several complementary decision procedures.
The main one is the \emph{stabilization-based}
% \paragraph{Stabilization-based procedure.}
%
% Our main decision procedure is based on the stabilization-based algorithm for
algorithm for
solving word equations with regular constraints introduced
in~\cite{NoodlerFM23} and later extended with efficient handling of length
constraints and disequations~\cite{NoodlerOOPSLA23}.
The stabilization-based algorithm starts, for every string variable, with
an~NFA encoding regular constraints on the variable and iteratively refines the
NFA according to the word equations until the stability condition is achieved.
The stability condition holds when, for every word equation, the language of
the left-hand side (obtained as the language of the concatenation of NFAs for
variables and string literals) equals the language of the right-hand side.
When stability is achieved, length constraints of the solutions are generated and passed to the LIA solver.
% \ol{}
% This tight cooperation between the feasible solutions and constraints given by equations and initial regular constraints 
% allows to significantly prune the generated state-space and avoid generating infeasible case splits. In order to handle length constraints as well, the stabilization-based 
% approach is extended with alignment and splitting of length variables (variables occurring in the length formula) in order to derive LIA formulae describing all 
% solutions for particular variable. The splitting is still steered by the feasible assignments maintaining the efficiency of the original stabilization algorithm. 
% The automata splitting uses extensions of classical non-deterministic automata constructions, implemented in \mata, to refine the languages until stability is achieved or one 
% of the languages becomes empty.
The algorithm is complete for the \emph{chain-free} \cite{ChainFree} combinations of equations,
regular and length constraints, together with unrestricted
disequations, making it the largest known decidable fragment of these types of
constraints.

% \paragraph{Handling of regular constraints.}
The stabilization-based decision procedure starts by inductively converting the
initial regular constraints into NFAs.
%
% Since the stabilization-based decision procedure represents languages of each variable in the form of NFAs, we convert the 
% initial regular constraints represented by regular expressions to the equivalent NFA form.
During the construction, we utilize eager simulation-based
reduction~\cite{bustan-sim-min,cece_foundation_2017} with on-demand
determinization and minimization.
In particular, if the regular expression contains complement, we try to keep
the automata from the corresponding regex subtrees in the form of minimal
deterministic automata.
% \lh{jenom sem zvedavy, tohle fakt je k necemu dobre?}

% \paragraph{Nielsen transformation.}
%
For an efficient handling of \emph{quadratic} equations (systems of equations
with at most two occurrences of each variable) with lengths, \noodler
implements a~decision procedure based on the \emph{Nielsen
transformation}~\cite{nielsen1917}.
The algorithm constructs a graph corresponding to the system and
reasons about it to determine if the input formula is
satisfiable or not~\cite{robson1999quadratic,ChenHLT23}.
If the system contains length variables, we also create a~counter automaton 
corresponding to the Nielsen graph (in a similar way as in~\cite{lin-quad-18}).
In the subsequent step, we contract edges, saturating the set of self-loops
and, finally, we iteratively generate flat counter sub-automata (a~flat
counter automaton only allows cycles that are self-loops), which are later
transformed into LIA formulae describing lengths of all possible
solutions (see an example in \cref{sec:nielsen-example}).  

% \paragraph{Regular language reasoning.}
%
In order to solve \emph{(dis)equations of regular expressions}, we reduce the
problem to reasoning about the corresponding NFAs (similarly as for regular
constraints handling).
In particular, we use efficient NFA equivalence and universality checking
from \mata, which implements advanced antichain-based
algorithms~\cite{dewulf_antichains_2006,abdulla_when_2010}.

%%%%%%%%%%%%%%%%%%%%%%%%%%%%%%%%%%%%%%%%%%%%%%%%%%%%%%%%%%%%%%%%%%
% \subsection{Preprocessing}
%%%%%%%%%%%%%%%%%%%%%%%%%%%%%%%%%%%%%%%%%%%%%%%%%%%%%%%%%%%%%%%%%%

\vspace{-2mm}
\paragraph{Preprocessing.}
Each decision procedure employs a sequence of preprocessing rules transforming
the string constraint to a more suitable form.
Our portfolio of rules includes transformations reducing the number of
equations by a conversion to regular constraints, propagating epsilons and
variables over equations, underapproximation rules, and rules reducing the
number of disequations (cf.~\cite{NoodlerOOPSLA23}).
On top of that, \ziiinoodler employs information about length-equivalent
variables allowing to infer simpler constraints (e.g., for $xy=zw$
with $\lenof{x} = \lenof{z}$, we can infer $y = w$).
\ziiinoodler also checks for simple unsatisfiable patterns for early termination.
A~sequence of preprocessing rules is composed 
for each of the decision procedures differently, maximizing their strengths.
%In particular, for 
%stabilization-based procedure, the rules aim at a transformation of (dis)equation to regular constraints \js{this sentence seem unnecessary}.

%%%%%%%%%%%%%%%%%%%%%%%%%%%%%%%%%%%%%%%%%%%%%%%%%%%%%%%%%%%%%%%%%%
% \subsection{Supported String Predicates and Limitations}
%%%%%%%%%%%%%%%%%%%%%%%%%%%%%%%%%%%%%%%%%%%%%%%%%%%%%%%%%%%%%%%%%%

\vspace{-2mm}
\paragraph{Supported String Predicates and Limitations.}
\ziiinoodler currently supports handling of basic string
predicates \replace, \substr, \at, \indexof, \prefix, \suffix, \contains, and a
limited support for $\neg\contains$.
%\js{should we mention also regular expression predicates? in\_re, to\_re, all predicates constructing regular languages}.
From the extended constraints, the core solver currently does not support the
\replaceall function (and variants of replacement based on regular expressions)
and \texttt{to}/\fromint conversions.
%\footnote{The strings-integers conversion is currently under development.}.
%\js{should we say that we can handle \isdigit and \tocode, \fromcode?} \vh{I wouldn't say that}.
The decision procedures used in \ziiinoodler make it complete for
the chain-free fragment with unbounded disequations and regular constraints~\cite{NoodlerOOPSLA23} (and for SAT assignments when if constraints are added), and quadratic equations.
% The decision procedures used in \ziiinoodler make it complete for SAT
% assignments falling into the chain-free fragment with unbounded disequations
% (together with length and regular constraints)~\cite{NoodlerOOPSLA23} and 
% quadratic equations.
% For quadratic equations with lengths we have an incomplete procedure.
Outside this fragment, our theory core is sound but incomplete.
% \lh{what about unsat chain-free}

%%%%%%%%%%%%%%%%%%%%%%%%%%%%%%%%%%%%%%%%%%%%%%%%%%%%%%%%%%%%%%%%%%
\vspace{-2.0mm}
\section{Experiments}
\vspace{-1.0mm}
%%%%%%%%%%%%%%%%%%%%%%%%%%%%%%%%%%%%%%%%%%%%%%%%%%%%%%%%%%%%%%%%%%

\newcommand{\tabResults}[0]{
\begin{table}[t]
\captionsetup{font=small}
%\vspace{-1mm}
\caption{
    Results of experiments on all benchmark sets.
    For each tool and benchmark set (as well as whole groups under
    $\Sigma$), we give the number of unsolved instances.
    Results for tools with the highest number of solved instances are in \textbf{bold}.
    Numbers with ${}^*$ contain also incorrect results.
}
\label{tab:results}
% \vspace{-3mm}
\hspace*{-4mm}
\resizebox{1.04\textwidth}{!}{%
\newcolumntype{g}{>{\columncolor{gray!30}}r}
\newcolumntype{f}{>{\columncolor{gray!30}}l}
\newcolumntype{h}{>{\columncolor{gray!30}}c}
\newcommand{\ping}[0]{\bf}
\newcommand{\wrongres}[2]{${}^*${#1}}
% if you want wrong results to be shown in parentheses, uncomment this
% \newcommand{\wrongres}[2]{${}^*${#1}({#2})}
\begin{tabular}{lgggggrrrrrrrrggggr}
\toprule
 & \multicolumn{5}{h}{\regexbench} & \multicolumn{8}{c}{\eqbench} & \multicolumn{4}{h}{\predbench} & \\\cmidrule(lr){2-6}\cmidrule(lr){7-14}\cmidrule(lr){15-18}
 & \multicolumn{1}{h}{\automatarkshort} & \multicolumn{1}{h}{\denghangshort} & \multicolumn{1}{h}{\stringfuzzshort} & \multicolumn{1}{h}{\sygusqgenshort} & \multicolumn{1}{h}{$\Sigma$} & \multicolumn{1}{c}{\kaluzashort} & \multicolumn{1}{c}{\keplerbenchshort} & \multicolumn{1}{c}{\nornbenchshort} & \multicolumn{1}{c}{\slentshort} & \multicolumn{1}{c}{\slogshort} & \multicolumn{1}{c}{\webappshort} & \multicolumn{1}{c}{\woorpjeshort} & \multicolumn{1}{c}{$\Sigma$} & \multicolumn{1}{h}{\fullstrintshort} & \multicolumn{1}{h}{\leetcodeshort} & \multicolumn{1}{h}{\strsmallshort} & \multicolumn{1}{h}{$\Sigma$} & \multicolumn{1}{c}{\pyexshort}\\
\emph{Included} & 15,995 & 999 & 10,050 & 343 & \sumcol{27,387} & 19,432 & 587 & 1,027 & 1,128 & 1,976 & 267 & 809 & \sumcol{25,226} & 11,669 & 2,652 & 1,670 & \sumcol{15,991} & 23,845\\
\emph{Unsupported} & 0 & 0 & 1,568 & 0 & \sumcol{1,568} & 0 & 0 & 0 & 0 & 0 & 414 & 0 & \sumcol{414} & 5,299 & 0 & 210 & \sumcol{5,509} & 0\\
\midrule
\rowcolor{GreenYellow}
\ziiinoodler       & 62       & \ping 0 & \ping 0             & \ping 0 & \ping \sumcol{62}            & 259               & \ping 4 & \ping 0 & \ping 5 & \ping 0 & \ping 0           & 243               & \sumcol{511}                 & 4       & 4       & 55                  & \sumcol{63}                & 4,424\\
\cvcv              & 94       & 18      & 1037                & \ping 0 & \sumcol{1149}                & \ping 0           & 240     & 85      & 22      & \ping 0 & 40                & 54                & \ping \sumcol{441}           & \ping 0 & \ping 0 & \ping 4             & \ping \sumcol{4}           & \ping 34\\
\ziii              & 113      & 118     & 340                 & \ping 0 & \sumcol{571}                 & 164               & 313     & 124     & 74      & 71      & 61                & \ping 25          & \sumcol{832}                 & 4       & \ping 0 & 32                  & \sumcol{36}                & 1,071\\
\ziiistriv         & 60       & 4       & 27                  & \ping 0 & \sumcol{91}                  & 174               & 254     & 73      & 73      & 16      & 62                & 78                & \sumcol{730}                 & 5       & 4       & 37                  & \sumcol{46}                & 570\\
\ostrich           & \ping 55 & 15      & 229                 & \ping 0 & \sumcol{299}                 & 288               & 387     & 1       & 130     & 7       & 65                & 53                & \sumcol{931}                 & 37      & 26      & \wrongres{106}{4}   & \sumcol{\wrongres{169}{0}} & 12,290\\
\ziiistriiire      & 66       & 27      & \wrongres{143}{50}  & 1       & \sumcol{\wrongres{237}{0}}   & \wrongres{144}{1} & 311     & 133     & 87      & 55      & \wrongres{104}{6} & \wrongres{118}{8} & \sumcol{\wrongres{952}{0}}   & 64      & 192     & \wrongres{179}{12}  & \sumcol{\wrongres{435}{0}} & 17,764\\
\ziiinoodleroopsla & 86       & 1       & \wrongres{1,014}{6} & \ping 0 & \sumcol{\wrongres{1,101}{0}} & 508               & 575     & \ping 0 & 6       & \ping 0 & \wrongres{3}{2}   & 256               & \sumcol{\wrongres{1,348}{0}} & 40      & 29      & \wrongres{493}{218} & \sumcol{\wrongres{562}{0}} & \wrongres{13,362}{6}\\
\bottomrule
\end{tabular}
%
% \vspace{-5mm}
}
\vspace{-4mm}
\end{table}
}

\vspace{-2mm}
%------------------------------------------------
\paragraph{Tools and environment.}
We compared \ziiinoodler with the following state-of-the-art tools:
\cvcv~\cite{cvc5} (version~1.0.8),
\ziii~\cite{z3} (version~4.12.2),
\ziiistriiire~\cite{Z3str3RE,BerzishDGKMMN23},
%\ziiitrau~\cite{abdulla_efficient_2020} (version~1.1),
\ziiistriv~\cite{MoraBKNG21}, 
\ostrich~\cite{AnthonyComplex2019}\footnote{Latest commit \texttt{70d01e2d2},
run with \texttt{-portfolio=strings} option.}, and \ziiinoodleroopsla
(version~0.1.0 used in~\cite{NoodlerOOPSLA23}).
We did not compare with \ziiitrau~\cite{Trau} as it is no longer under active development
and gives incorrect results on newer benchmarks.
% We also add the version of \ziiinoodler , labelled
% as \ziiinoodleroopsla, for comparison.
%
The experiments were executed on a~workstation with an Intel Xeon Silver 4314 CPU @ 2.4\,GHz with 128\,GiB of RAM
running Debian GNU/Linux.
The timeout was set to 120\,s, memory limit was set to 8\,GiB.

\vspace{-3mm}
%------------------------------------------------
\paragraph{Benchmarks.}
The benchmarks come from the SMT-LIB~\cite{SMTLIB} repository, specifically
categories \QFS~\cite{QF_S} and \QFSLIA~\cite{QF_SLIA}.
These benchmarks were also used in SMT-COMP'23~\cite{SMTCOMP23}, in which
\ziiinoodler participated (version 0.2.0).
As \ziiinoodler does not support \texttt{to}/\fromint conversions and \replaceall-like
predicates, we excluded formulae whose satisfiability checking needs their
support.
% Moreover, we also removed instances easily solvable by a specific algorithm
% (see below).
%
Based on the occurrences of different kinds of constraints, we
divide the benchmarks into three groups:
\begin{description}
    \item[\regexbench] This category contains formulae with dominating regular
      membership and length constraints. 
      It consists of \automatark~\cite{Z3str3RE}, \denghang,
      \stringfuzz~\cite{StringFuzz}, and \sygusqgen benchmark sets.
      We excluded 1,568 formulae from \stringfuzz that require support of
      the \toint predicate.
    
    % contains 1568 formulae 
    % which are not supported by \ziiinoodler (\toint predicate), therefore we do not include them in the evaluation.
    
    \item[\eqbench] The formulae in this category consist mostly of word
      equations with length constraints and a~small amount of other predicates.
    It contains \kaluza~\cite{saxena2020kaluza,tinelli-fmsd16}, \keplerbench~\cite{LeH18}, \nornbench~\cite{AutomataSplitting,Norn}, \slent~\cite{slent}, \slog~\cite{fang-yu-circuits}, \webapp, and \woorpje~\cite{DayEKMNP19} benchmark sets.
    We excluded 414 formulae from \webapp that require support of \replaceall, \replacere, and \replacereall predicates.
    %We again do not include formulae unsupported by \ziiinoodler, more specifically 414 formulae from \webapp (they contain \replaceall, \replacere, and \replacereall).
    % We also omitted \woorpje benchmark from SMT-LIB as it can be easily solved by a specialized algorithm~\cite{DayEKMNP19}. 

    % Furthermore, this category should also contain \woorpje~\cite{DayEKMNP19} benchmark set, but it is quite different from other benchmarks, the equations are not \emph{chain-free} nor \emph{quadratic}, therefore they are especially hard for \ziiinoodler. \js{maybe some other bullshit about \woorpje, that it is bullshit benchmark and there is decision procedure that we will implement.}
    
    \item[\predbench] Although \ziiinoodler focuses mainly on word equations
      with length and regular constraints, the evaluation includes also a group consisting
      of smaller formulae that use string predicates such as
      \substr, \at, \contains, etc.
    It is formed from \fullstrint, \leetcode, and \strsmall~\cite{StrSmallRw} benchmark sets.
    We removed 5,509 formulae containing the \texttt{to}/\fromint predicates from \fullstrint and \strsmall.

    %We should also include two other benchmark sets in this category: \transducerplus~\cite{AnthonyInteger2020} and \pyex~\cite{cvc417}.
    %\transducerplus is a small benchmark (91 formulae) that is not supported by \ziiinoodler, because it contains \replaceall predicate.
    %On the other hand, \pyex is a really large benchmark that would dominate this category. %, we therefore give results for it separately.
    %\js{some better bullshit about \pyex?}
\end{description}
\vspace{-1mm}
We also consider the \pyex~\cite{cvc417} benchmark, which we do not put
into any of these groups, as it contains large formulae with complex predicates (\substr, \contains, etc.). %\lh{what do you mean by predicates? Extended constraints? My impression is that the word predicates was not introduced in this meaning.}
We note that we omit the small \transducerplus~\cite{AnthonyInteger2020}
benchmark because it contains exclusively formulae with \replaceall.

\begin{table}[t]
\captionsetup{font=small}
%\vspace{-1mm}
\caption{
    Results of experiments on all benchmark sets.
    For each tool and benchmark set (as well as whole groups under
    $\Sigma$), we give the number of unsolved instances.
    Results for tools with the highest number of solved instances are in \textbf{bold}.
    Numbers with ${}^*$ contain also incorrect results.
}
\label{tab:results}
% \vspace{-3mm}
\hspace*{-4mm}
\resizebox{1.04\textwidth}{!}{%
\newcolumntype{g}{>{\columncolor{gray!30}}r}
\newcolumntype{f}{>{\columncolor{gray!30}}l}
\newcolumntype{h}{>{\columncolor{gray!30}}c}
\newcommand{\ping}[0]{\bf}
\newcommand{\wrongres}[2]{${}^*${#1}}
% if you want wrong results to be shown in parentheses, uncomment this
% \newcommand{\wrongres}[2]{${}^*${#1}({#2})}
\begin{tabular}{lgggggrrrrrrrrggggr}
\toprule
 & \multicolumn{5}{h}{\regexbench} & \multicolumn{8}{c}{\eqbench} & \multicolumn{4}{h}{\predbench} & \\\cmidrule(lr){2-6}\cmidrule(lr){7-14}\cmidrule(lr){15-18}
 & \multicolumn{1}{h}{\automatarkshort} & \multicolumn{1}{h}{\denghangshort} & \multicolumn{1}{h}{\stringfuzzshort} & \multicolumn{1}{h}{\sygusqgenshort} & \multicolumn{1}{h}{$\Sigma$} & \multicolumn{1}{c}{\kaluzashort} & \multicolumn{1}{c}{\keplerbenchshort} & \multicolumn{1}{c}{\nornbenchshort} & \multicolumn{1}{c}{\slentshort} & \multicolumn{1}{c}{\slogshort} & \multicolumn{1}{c}{\webappshort} & \multicolumn{1}{c}{\woorpjeshort} & \multicolumn{1}{c}{$\Sigma$} & \multicolumn{1}{h}{\fullstrintshort} & \multicolumn{1}{h}{\leetcodeshort} & \multicolumn{1}{h}{\strsmallshort} & \multicolumn{1}{h}{$\Sigma$} & \multicolumn{1}{c}{\pyexshort}\\
\emph{Included} & 15,995 & 999 & 10,050 & 343 & \sumcol{27,387} & 19,432 & 587 & 1,027 & 1,128 & 1,976 & 267 & 809 & \sumcol{25,226} & 11,669 & 2,652 & 1,670 & \sumcol{15,991} & 23,845\\
\emph{Unsupported} & 0 & 0 & 1,568 & 0 & \sumcol{1,568} & 0 & 0 & 0 & 0 & 0 & 414 & 0 & \sumcol{414} & 5,299 & 0 & 210 & \sumcol{5,509} & 0\\
\midrule
\rowcolor{GreenYellow}
\ziiinoodler       & 62       & \ping 0 & \ping 0             & \ping 0 & \ping \sumcol{62}            & 259               & \ping 4 & \ping 0 & \ping 5 & \ping 0 & \ping 0           & 243               & \sumcol{511}                 & 4       & 4       & 55                  & \sumcol{63}                & 4,424\\
\cvcv              & 94       & 18      & 1037                & \ping 0 & \sumcol{1149}                & \ping 0           & 240     & 85      & 22      & \ping 0 & 40                & 54                & \ping \sumcol{441}           & \ping 0 & \ping 0 & \ping 4             & \ping \sumcol{4}           & \ping 34\\
\ziii              & 113      & 118     & 340                 & \ping 0 & \sumcol{571}                 & 164               & 313     & 124     & 74      & 71      & 61                & \ping 25          & \sumcol{832}                 & 4       & \ping 0 & 32                  & \sumcol{36}                & 1,071\\
\ziiistriv         & 60       & 4       & 27                  & \ping 0 & \sumcol{91}                  & 174               & 254     & 73      & 73      & 16      & 62                & 78                & \sumcol{730}                 & 5       & 4       & 37                  & \sumcol{46}                & 570\\
\ostrich           & \ping 55 & 15      & 229                 & \ping 0 & \sumcol{299}                 & 288               & 387     & 1       & 130     & 7       & 65                & 53                & \sumcol{931}                 & 37      & 26      & \wrongres{106}{4}   & \sumcol{\wrongres{169}{0}} & 12,290\\
\ziiistriiire      & 66       & 27      & \wrongres{143}{50}  & 1       & \sumcol{\wrongres{237}{0}}   & \wrongres{144}{1} & 311     & 133     & 87      & 55      & \wrongres{104}{6} & \wrongres{118}{8} & \sumcol{\wrongres{952}{0}}   & 64      & 192     & \wrongres{179}{12}  & \sumcol{\wrongres{435}{0}} & 17,764\\
\ziiinoodleroopsla & 86       & 1       & \wrongres{1,014}{6} & \ping 0 & \sumcol{\wrongres{1,101}{0}} & 508               & 575     & \ping 0 & 6       & \ping 0 & \wrongres{3}{2}   & 256               & \sumcol{\wrongres{1,348}{0}} & 40      & 29      & \wrongres{493}{218} & \sumcol{\wrongres{562}{0}} & \wrongres{13,362}{6}\\
\bottomrule
\end{tabular}
%
% \vspace{-5mm}
}
\vspace{-4mm}
\end{table}
 %%%%%%%%%%%%%%%%%%%%%%%%%%%%%

\newcommand{\figScatter}[0]{
\begin{figure}[tb]%
\centering
\begin{subfigure}{0.33\textwidth}
    \centering
    \includegraphics[width=\linewidth]{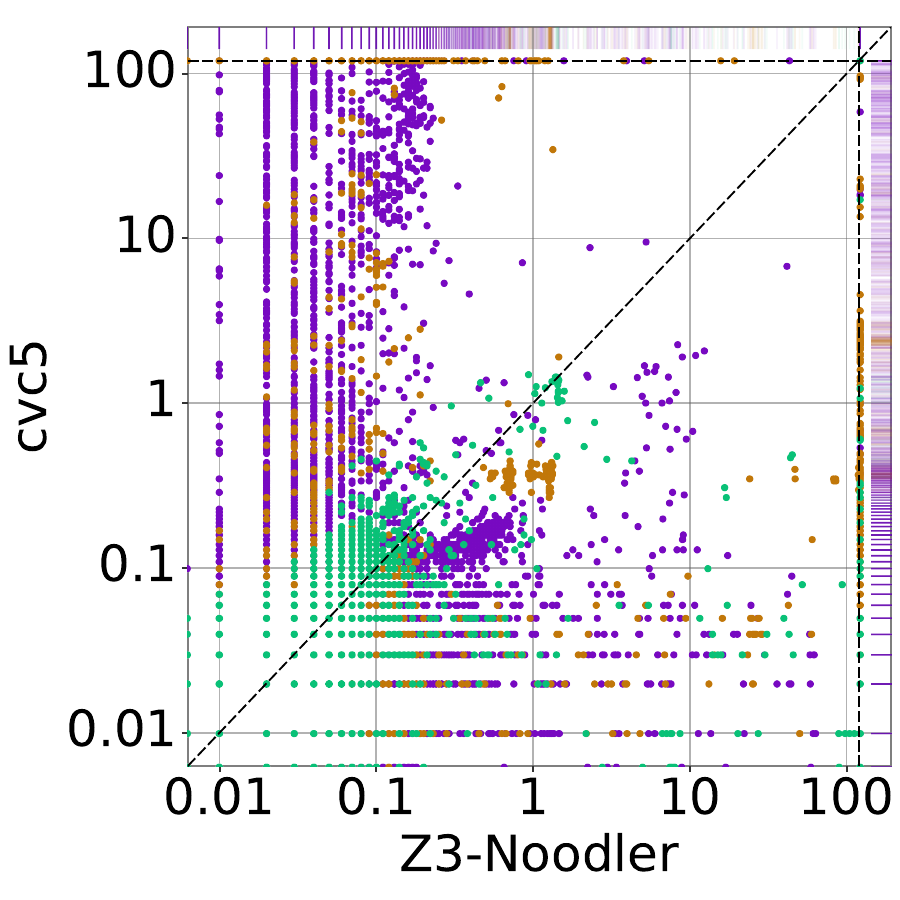}
    \vspace{-6mm}
    \caption{\ziiinoodler vs. \cvcv}
    \label{fig:cvc5}
\end{subfigure}\hfil
\begin{subfigure}{0.33\textwidth}
    \centering
    \includegraphics[width=\linewidth]{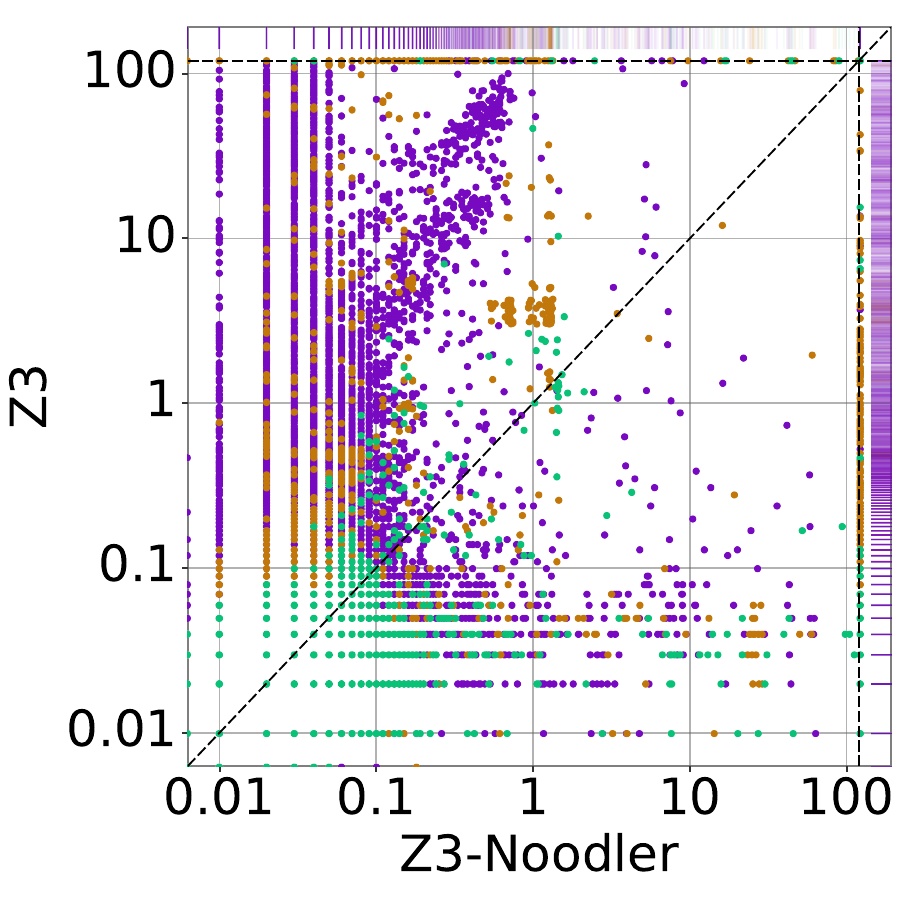}
    \vspace{-6mm}
    \caption{\ziiinoodler vs. \ziii}
    \label{fig:z3}
\end{subfigure}\hfil
\begin{subfigure}{0.33\textwidth}
    \centering
    \includegraphics[width=\linewidth]{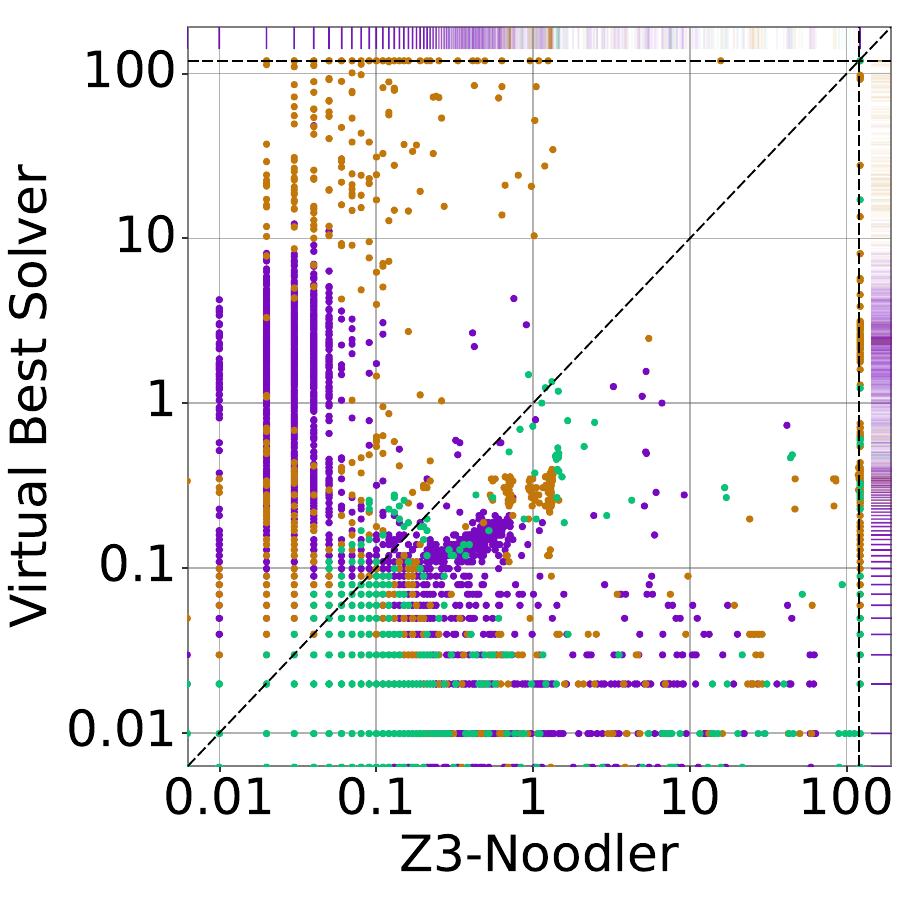}
    \vspace{-6mm}
    \caption{\ziiinoodler vs. VBS}
    \label{fig:z3str4}
\end{subfigure}
% \begin{subfigure}{0.33\textwidth}
%     \centering
%     \includegraphics[width=\linewidth]{figs/noodler_vs_ostrich.pdf}
%     \caption{\ziiinoodler vs. \ostrich}
%     \label{fig:ostrich}
% \end{subfigure}\hfil
% \begin{subfigure}{0.33\textwidth}
%     \centering
%     \includegraphics[width=\linewidth]{figs/noodler_vs_z3str3re.pdf}
%     \caption{\ziiinoodler vs. \ziiistriiire}
%     \label{fig:z3str3RE}
% \end{subfigure}\hfil
% \begin{subfigure}{0.33\textwidth}
%     \centering
%     \includegraphics[width=\linewidth]{figs/noodler_vs_old_noodler.pdf}
%     \caption{\ziiinoodler vs. \ziiinoodleroopsla}
%     \label{fig:oldnoodler}
% \end{subfigure}
\vspace{-2mm}
\caption{
    Comparison of \ziiinoodler with \cvcv, \ziii, and the virtual best solver (VBS).
    Times are in seconds, axes are logarithmic.
    Dashed lines represent timeouts (120\,s).
    Colours distinguish groups:
    \RGBcircle{120,10,195}\,\regexbench,
    \RGBcircle{194,119,9}\,\eqbench, and
    \RGBcircle{36,194,120}\,\predbench.
}
\label{fig:scatter}
\vspace*{-6mm}
\end{figure}
}

%!!!!!!!!!!!!!!!!!!!!!!!!!!!!!!!!
\enlargethispage{2mm}
%!!!!!!!!!!!!!!!!!!!!!!!!!!!!!!!!
%----------------------------------------
\vspace{-3mm}
\paragraph{Results.}
\begin{wraptable}[10]{r}{4.5cm} % 12 is number of lines that are wrapped
\captionsetup{font=small}
\vspace*{-8mm}
\caption{Average run times (in seconds) of solved instances and their standard
  deviations.}
\label{tab:average}
\vspace{-2mm}
\hspace*{-2mm}
\resizebox{1.04\linewidth}{!}{%
\newcolumntype{g}{>{\columncolor{gray!30}}r}
\newcolumntype{f}{>{\columncolor{gray!30}}l}
\newcolumntype{h}{>{\columncolor{gray!30}}c}
\newcommand{\ping}[0]{\bf}
\begin{tabular}{lggrrgg}
\toprule
& \multicolumn{2}{h}{\bf{Reg}} & \multicolumn{2}{c}{\bf{Eq}} & \multicolumn{2}{h}{\bf{Pred}}\\
& \multicolumn{1}{h}{avg} & \multicolumn{1}{h}{std} & \multicolumn{1}{c}{avg} & \multicolumn{1}{c}{std} & \multicolumn{1}{h}{avg} & \multicolumn{1}{h}{std}\\
\midrule
\rowcolor{GreenYellow}
\ziiinoodler & \ping 0.11 & 1.35 &\ping 0.11 & 2.13 & 0.11 & 2.16 \\
\cvcv & 1.17 & 8.51 & \ping 0.11 & 2.15 & 0.03 & 0.15 \\
\ziii & 1.92 & 9.71 & 0.18 & 2.83 & 0.04 & 0.42 \\
\ziiistriv & 0.35 & 2.00 & 0.25 & 3.40 & 0.02 & 0.31 \\
\ostrich & 4.29 & 8.67 & 4.28 & 9.28 & 12.71 & 15.08 \\
\ziiistriiire & 0.31 & 3.28 & 0.13 & 2.72 & \ping 0.01 & 0.08 \\
\ziiinoodleroopsla & 0.27 & 2.86 & 0.12 & 2.93 & 0.09 & 1.69 \\
% \vbs & 0.11 & 0.60 & 0.27 & 3.97 & 0.01 & 0.14 \\
\bottomrule
\end{tabular}
}
\end{wraptable}
We show the number of unsolved instances for each benchmark and tool (as well as whole groups) in
\cref{tab:results}.
Some tools gave incorrect results (determined by comparing to the output of \cvcv and~\ziii)
for some benchmarks.
Usually, this was less than 10 instances, except for \ziiistriiire on \stringfuzz
and \strsmall (50 and 12 incorrect results respectively) and \ziiinoodleroopsla
on \strsmall (218 incorrect results).
\Cref{tab:average} then shows the average run times and their standard deviations for solved instances for each category and tool. 
%while
%the scatter plot comparing runtimes of \ziiinoodler with other tools is shown in \cref{fig:scatter}. 

The results show that \ziiinoodler outperforms other tools on the \regexbench group
(in particular on \denghang, \stringfuzz, and \sygusqgen) both in the number of
solved instances and the average run time.
Only on \automatark it 
cannot solve the most formulae (but it solves only 7 less than the winner \ostrich, while being much faster).
%\lh{mention much smaller runtime, perhaps even in concrete numbers}).
% \ziiinoodler has also the lowest average time compared to the other competitors. 

On the \eqbench group, \ziiinoodler also outperforms other tools on most of the benchmarks. 
In particular on \keplerbench, \nornbench, \slent, \slog, and \webapp.
On \kaluza, it is outperformed by other tools, but it still solves the vast majority of formulae.
\ziiinoodler has worse performance on \woorpje, which seems to be a synthetic benchmark generated to showcase the strength of a~specialized algorithm~\cite{DayEKMNP19}
(this benchmark is the reason for \ziiinoodler taking the second place in the
whole group).
With 0.11\,s, \ziiinoodler and \cvcv have the lowest average run time.

The winner of \predbench is \cvcv.
% On \predbench, \ziiinoodler is outperformed by \cvcv in the number of solved instances. 
In particular, on \fullstrint and \leetcode the difference with \ziiinoodler is
equally 4 instances and on \strsmall the difference is 51 cases. The average
time of \ziiinoodler is also a bit higher, with 0.11\,s for \ziiinoodler
compared to the 0.03\,s for \cvcv.
Similarly, \ziiinoodler is outperformed by \cvcv, \ziii, and \ziiistriv on \pyex.
Indeed, we have not optimized \ziiinoodler for formulae with large numbers of predicates yet.
The results of \ziiinoodler could, however, be further improved 
by proper axiom saturation for predicates or lazy predicate evaluation.
% (we leave this as future work).

\begin{wraptable}[9]{r}{4.9cm}
\vspace*{-8mm}
\captionsetup{font=small}
% \caption{Virtual best solvers in different configurations, showing total number of unsolved instances and total time (for solved instances) in seconds.}
\caption{Evaluating solver contribution to a portfolio. Times are in seconds.}
\label{tab:vbs}
\vspace{-3mm}
\hspace*{-3mm}
\resizebox{1.04\linewidth}{!}{%
\newcolumntype{g}{>{\columncolor{gray!30}}r}
\newcolumntype{f}{>{\columncolor{gray!30}}l}
\newcolumntype{h}{>{\columncolor{gray!30}}c}
\newcommand{\ping}[0]{\bf}
\begin{tabular}{lggrr}
\toprule
& \multicolumn{2}{h}{\regexbench} & \multicolumn{2}{c}{\eqbench} \\
& \multicolumn{1}{h}{Unsolved} & \multicolumn{1}{h}{Time} & \multicolumn{1}{c}{Unsolved} & \multicolumn{1}{c}{~~Time~~}\\
\midrule
\rowcolor{GreenYellow}
\vbsnoodler                         & 1   & 427    & 19  & 1,304 \\
\vbsnoodler - \ziiinoodler          & 1   & 2,914   & 131 & 6,830 \\
\vbsnoodler - \cvcv                 & 1   & 549    & 145 & 1,401 \\
\vbsnoodler - \ziii                 & 1   & 430    & 29  & 1,579 \\
\vbsnoodler - \ziiistriv            & 1   & 473    & 19  & 1,416 \\
\vbsnoodler - \ostrich              & 1   & 427    & 21  & 1,270 \\
\vbsnoodler - \ziiistriiire         & 1   & 510    & 20  & 1,307 \\
\cvcv + \ziii + \ziiinoodler & 1   & 608    & 22  & 1,471 \\
\cvcv + \ziii                & 278 & 27,916  & 303 & 2,805 \\
\bottomrule
\end{tabular}
}
\end{wraptable}

\figScatter %%%%%%%%%%%%%%%%%%%%%%%%%%%%%%%%%%%%%%%%%

In \cref{fig:scatter} we show scatter plots comparing running time of
\ziiinoodler with \cvcv, \ziii, and virtual best solver (VBS; a~solver that
takes the best result from all tools other than \ziiinoodler) on all three
benchmark groups. 
% We do not give comparison with \ostrich, as it is significantly slower than other solvers nor do we show comparison with \ziiistriiire, as it has too many 
% incorrect results (their plots are shown in \cref{sec:detailed-results}).
See~\cref{sec:detailed-results} for comparison with other solvers.
The plots show that \ziiinoodler outperforms the competitors on a~vast number
of instances, in many cases being complementary to them.
To validate this claim, we also checked how different solvers contribute to a~portfolio.
That is, we took the VBS including \ziiinoodler (\vbsnoodler) and then checked
how well the portfolio works without each of the solvers.
\Cref{tab:vbs} shows the results on the \regexbench and \eqbench groups (we
omit \predbench, where \ziiinoodler does not help the portfolio).
The results show that on the two groups, \ziiinoodler is the most valuable
solver in the portfolio.
We also include results on the small portfolio of \ziii and \cvcv (with and
without \ziiinoodler) showing that, on the two groups, using just these three
solvers is almost as good as using the whole portfolio of all solvers.

% The second observation is that \ziiinoodler is complementary to the rest of the
% tools, which might be beneficial especially for its use in portfolio solvers.
% \Cref{tab:vbs} also shows this
% \js{this table will probably not be here, we can maybe just tell the numbers in the text}.
% \ol{deal with this}

% See~\cref{sec:detailed-results} for comparison with other solvers.
% The plots show that \ziiinoodler outperforms the competitors on a~vast number
% of instances.
% The second observation is that \ziiinoodler is complementary to the rest of the
% tools, which might be beneficial especially for its use in portfolio solvers.
% \Cref{tab:vbs} also shows this
% \js{this table will probably not be here, we can maybe just tell the numbers in the text}.
% \ol{deal with this}

%!!!!!!!!!!!!!!!!!!!!!!!!!!!!!!!!
\enlargethispage{2mm}
%!!!!!!!!!!!!!!!!!!!!!!!!!!!!!!!!

Comparing with the older version \ziiinoodleroopsla from \cite{NoodlerOOPSLA23}, we can see that there is a~significant improvement in most benchmarks, most significantly in \automatark, \stringfuzz, \keplerbench, \strsmall, and \kaluza.
We note that adding more complicated algorithm selection strategies
significantly improved the overall performance of \ziiinoodler, but, on the other
hand, decreased the performance on \kaluza (cf.\
\cite{NoodlerOOPSLA23}).
%
% Note that in~\cite{NoodlerOOPSLA23}, \ziiinoodleroopsla had better results on \kaluza than is presented here,
% because it was run with an
% underapproximation heuristic on the whole benchmark \lh{what is "whole benchmark" Why is the heuristic off now if it is good?}.
% For the comparison here, we used the default behaviour.
Better results in \automatark and \stringfuzz stem from the improvements in
\mata and from heuristics tailored for regular expressions handling.
Including Nielsen's algorithm~\cite{nielsen1917} has the largest impact on the
\keplerbench benchmark.
The improvement on predicate-intensive benchmarks is caused by optimizations in
axiom saturation for predicates.
The older version also had multiple bugs that have been fixed in the current version.

\bibliography{literature.bib}

\clearpage
\appendix

\section{Logic of Strings}\label{sec:string-logic}
Let $\Sigma$ be a finite alphabet. We use $\Sigma^*$ to denote the set of 
all finite words over~$\Sigma$.
% The empty word, we denote as $\epsilon$.
In \ziiinoodler, we consider the logic of strings
over~$\Sigma$, a set of string variables $\varss$ ranging over $\Sigma^*$, and
a~set of integer variables $\varsi$ ranging over $\mathbb{Z}$, with the syntax
of a~formula~$\varphi$ given using the following grammar:
\begin{align*} 
    \varphi &:= \varphi_i \mid \varphi_s \mid \varphi \wedge \varphi \mid \neg\varphi  \\
    \varphi_s &:= t_s = t_s \mid t_s \in \regex \mid \regex = \regex \mid \contains(t_s, t_s) \mid \prefix(t_s, t_s) \mid \suffix(t_s, t_s)\\% \mid \isdigit(t_s)\\
    \varphi_i &:=  t_i \leq t_i  \\
    t_s &:=  v_s \mid w \mid t_s \cdot t_s \mid \substr(t_s, t_i, t_i) \mid \at(t_s, t_i) \mid \replace(t_s, t_s, t_s)\\% \mid \fromcode(t_i) \\%\mid \fromint(t_i) \\ 
    t_i &:= v_i \mid k \mid t_i \cdot t_i \mid \lenof{t_s} \mid \indexof(t_s, t_s)% \mid \tocode(t_s) %\mid \toint(t_s)
\end{align*}
with $\regex$ is a~regular language, $v_s \in \varss$, $w \in \Sigma^*$, $v_i
\in \varsi$, and $k \in \mathbb{Z}$.
Here, a~string atomic formula $\varphi_s$ is given as either an equation of string
terms $t_s$, a regular constraint checking whether 
the value of a~string term belongs to a~regular language (specified by
an extended regular expression\footnote{I.e., regular expressions enriched
with operators for intersection and complement.}), regex equality, or
application of predicates $\contains$, $\prefix$, and $\suffix$. %, and $\isdigit$.
String terms are then concatenations of strings and variables extended by
string predicates $\substr$, $\at$, and $\replace$. %and $\fromcode$.
A~length atomic formula is given as a~\emph{linear integer arithmetic} atomic
formula with terms containing also $\lenof{t_s}$ denoting the length of~$t_s$
and application of $\indexof$. % or $\tocode$.
We consider the usual semantics of the predicates as defined by
\smtlib~\cite{SMTLIB-Strings}.

%%%%%%%%%%%%%%%%%%%%%%%%%%%%%%%%%%%%%%%%%%%%%%%%%%%%%%%%%%%%%%%%%%
\section{Example of Nielsen Transformation}\label{sec:nielsen-example}
%%%%%%%%%%%%%%%%%%%%%%%%%%%%%%%%%%%%%%%%%%%%%%%%%%%%%%%%%%%%%%%%%%

\begin{figure}[t]
    \begin{subfigure}[b]{0.4\textwidth}
        \centering
        \begin{tikzpicture}[
    auto,
    transform shape,
    node distance=3cm,->,>=stealth,
    scale=0.8
  ]
  
  \tikzstyle{state} = [rectangle,draw,rounded corners=.1cm,draw=black,minimum height=0.7cm, inner sep=3pt]

  \node[state, initial, initial text={}] (s0) at (0,0) {$xy = ax$};
  \node[state, right of=s0] (s2) {$y = a$};
  \node[state, below of=s2, node distance=1.5cm] (s4) {$y = \epsilon$};
  \node[state, left of=s4] (s5) {$\epsilon = \epsilon$};

  \draw (s0) edge [loop above] node[above] {${x}\hookrightarrow{ax}$} (s0);
  \draw (s0) to node[above] {${x}\hookrightarrow{\epsilon}$} (s2);
  \draw (s2) to node[left] {${y}\hookrightarrow{ay}$} (s4);
  \draw (s4) to node[above] {${y}\hookrightarrow{\epsilon}$} (s5);
  
  \end{tikzpicture}
  
        \caption{}
        \label{fig:nielsen}
    \end{subfigure}
    ~
    \begin{subfigure}[b]{0.58\textwidth}
        \centering
        \begin{tikzpicture}[
    auto,
    transform shape,
    node distance=2cm,->,>=stealth,
    scale=0.8
  ]
  
  \tikzstyle{state} = [circle,draw,draw=black,minimum height=0.5cm, inner sep=2pt]

  \node[state,initial,initial text={}] (s5) {};
  \node[state, right of=s5] (s4) {};
  \node[state, right of=s4, node distance=2.5cm] (s2) {};
  \node[state, right of=s2, initial, initial text={},accepting] (s0) {};

  \draw (s0) edge [loop above] node[above] {$x := x + 1$} (s0);
  \draw (s2) to node[above] {$x := 0$} (s0);
  \draw (s4) to node[above] {$y := y + 1$} (s2);
  \draw (s5) to node[above] {$y := 0$} (s4);
  
  \end{tikzpicture}
  
        \caption{}
        \label{fig:counter-system}
    \end{subfigure}

    \caption{
        Example of Nielsen graph and the corresponding counter system.
    }
    \label{fig:quad}
\end{figure}
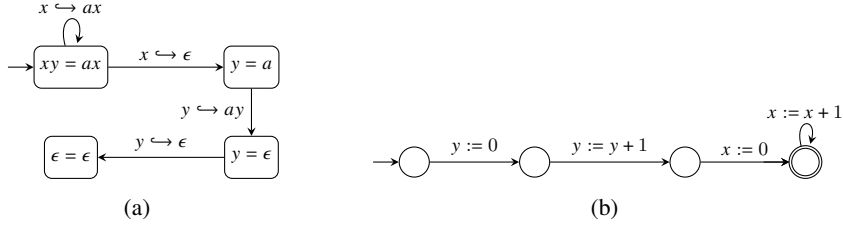

Consider the quadratic equation $xy=ax$ with the corresponding Nielsen graph \cref{fig:nielsen} (redundant nodes are omitted). The Nielsen graph is 
        derived from the initial equation $yx=ax$ by application of rewriting rules and trimming the common prefix. For instance if we apply the substitution 
        ${y}\hookrightarrow{ay}$ on the node $y = a$, we get $ay = a$ yielding to $y = \epsilon$.
        The counter system corresponding to the Nielsen graph is shown in \cref{fig:counter-system}. 
        Since the system is flat, the length formula describing lengths of $x$ and $y$ is given as 
        $\varphi(x,y) = \exists y_0, y_1, x_0, x_1: y_0 = 0 \wedge y_1 = y_0 + 1 \wedge x_0 = 0 \wedge (\exists k: k \geq 0 \wedge x_1 = x_0 + k) \wedge x_1 = x \wedge y_1 = y$, 
        which is equivalent to $y=1 \wedge x \geq 0$.

%%%%%%%%%%%%%%%%%%%%%%%%%%%%%%%%%%%%%%%%%%%%%%%%%%%%%%%%%%%%%%%%%%
\section{Detailed Results}\label{sec:detailed-results}
%%%%%%%%%%%%%%%%%%%%%%%%%%%%%%%%%%%%%%%%%%%%%%%%%%%%%%%%%%%%%%%%%%

\begin{figure}[h]%
  \captionsetup{font=small}
\centering
\begin{subfigure}{0.33\textwidth}
    \centering
    \includegraphics[width=\linewidth]{figs/noodler_vs_cvc5.pdf}
    \caption{\ziiinoodler vs. \cvcv}
    \label{fig:cvc5}
\end{subfigure}\hfil
\begin{subfigure}{0.33\textwidth}
    \centering
    \includegraphics[width=\linewidth]{figs/noodler_vs_z3.pdf}
    \caption{\ziiinoodler vs. \ziii}
    \label{fig:z3}
\end{subfigure}\hfil
\begin{subfigure}{0.33\textwidth}
    \centering
    \includegraphics[width=\linewidth]{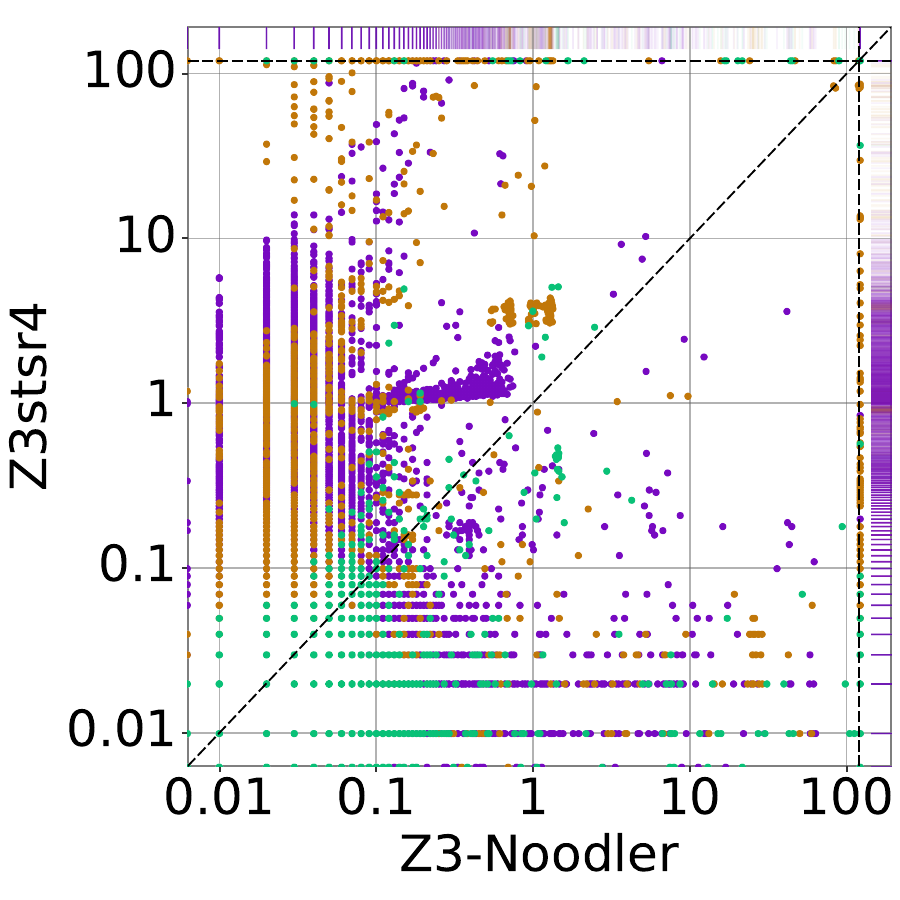}
    \caption{\ziiinoodler vs. \ziiistriv}
    \label{fig:z3str4}
\end{subfigure}

\begin{subfigure}{0.33\textwidth}
    \centering
    \includegraphics[width=\linewidth]{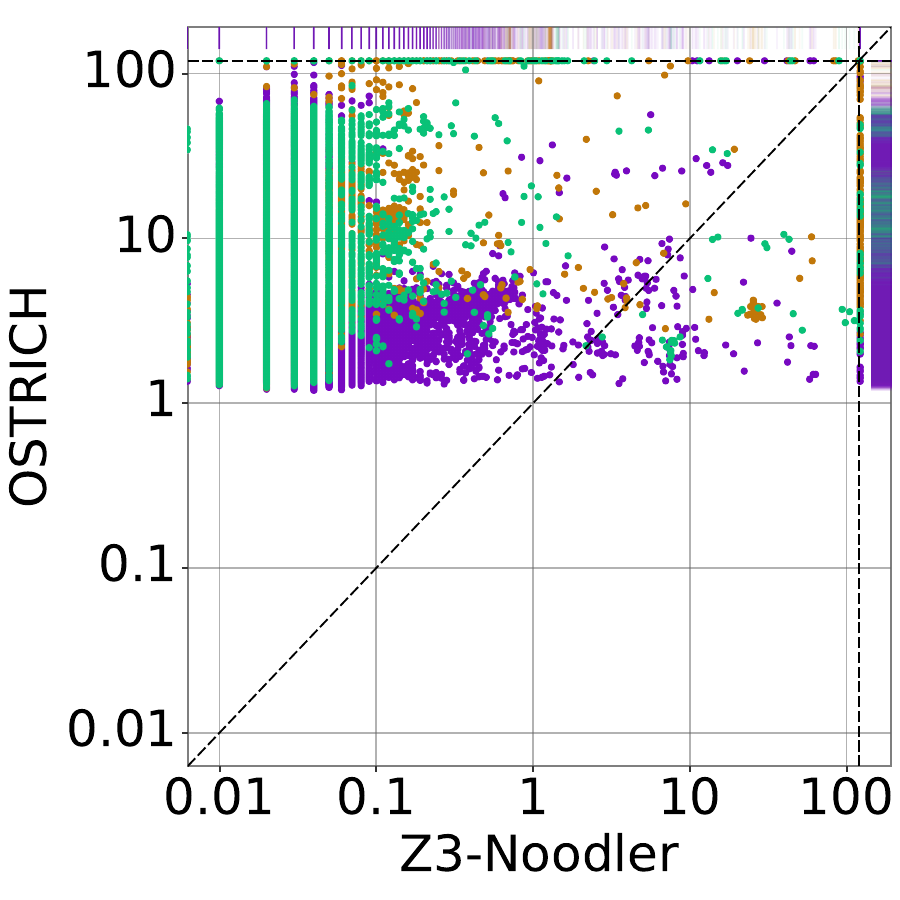}
    \caption{\ziiinoodler vs. \ostrich}
    \label{fig:ostrich}
\end{subfigure}\hfil
\begin{subfigure}{0.33\textwidth}
    \centering
    \includegraphics[width=\linewidth]{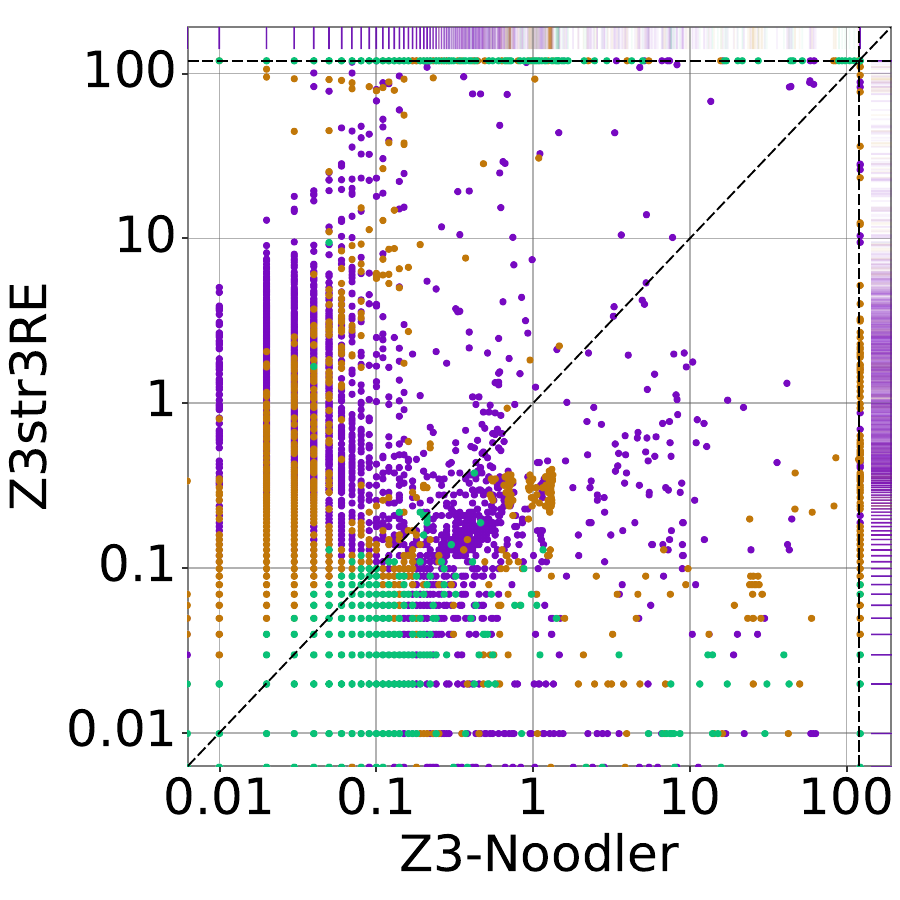}
    \caption{\ziiinoodler vs. \ziiistriiire}
    \label{fig:z3str3RE}
\end{subfigure}\hfil
\begin{subfigure}{0.33\textwidth}
    \centering
    \includegraphics[width=\linewidth]{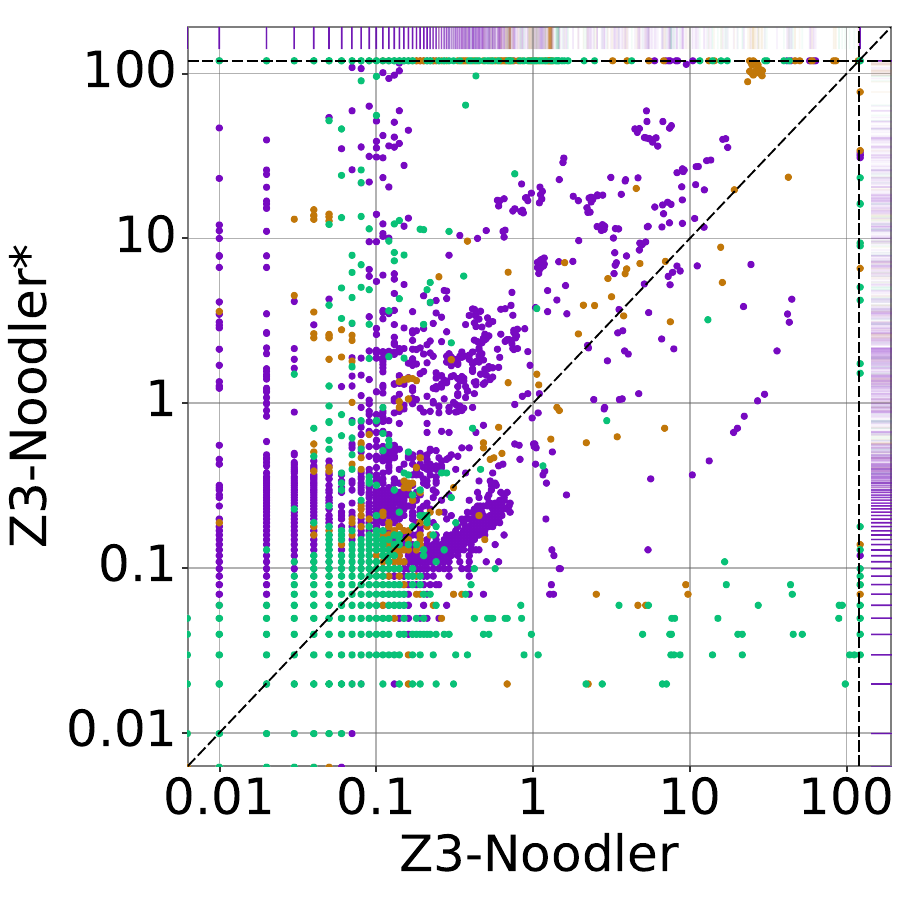}
    \caption{\ziiinoodler vs. \ziiinoodleroopsla}
    \label{fig:oldnoodler}
\end{subfigure}
\caption{
    Comparison of \ziiinoodler with other tools on all three benchmark categories.
    Times are in seconds, axes are logarithmic.
    Dashed lines represent timeouts (120\,s).
    Colors distinguish categories:
    \RGBcircle{120,10,195}\,\regexbench,
    \RGBcircle{194,119,9}\,\eqbench, and
    \RGBcircle{36,194,120}\,\predbench.
}
\label{fig:scatter-all}
\end{figure}

\begin{figure}[h]%
  \captionsetup{font=small}
\centering
\begin{subfigure}{0.33\textwidth}
    \centering
    \includegraphics[width=\linewidth]{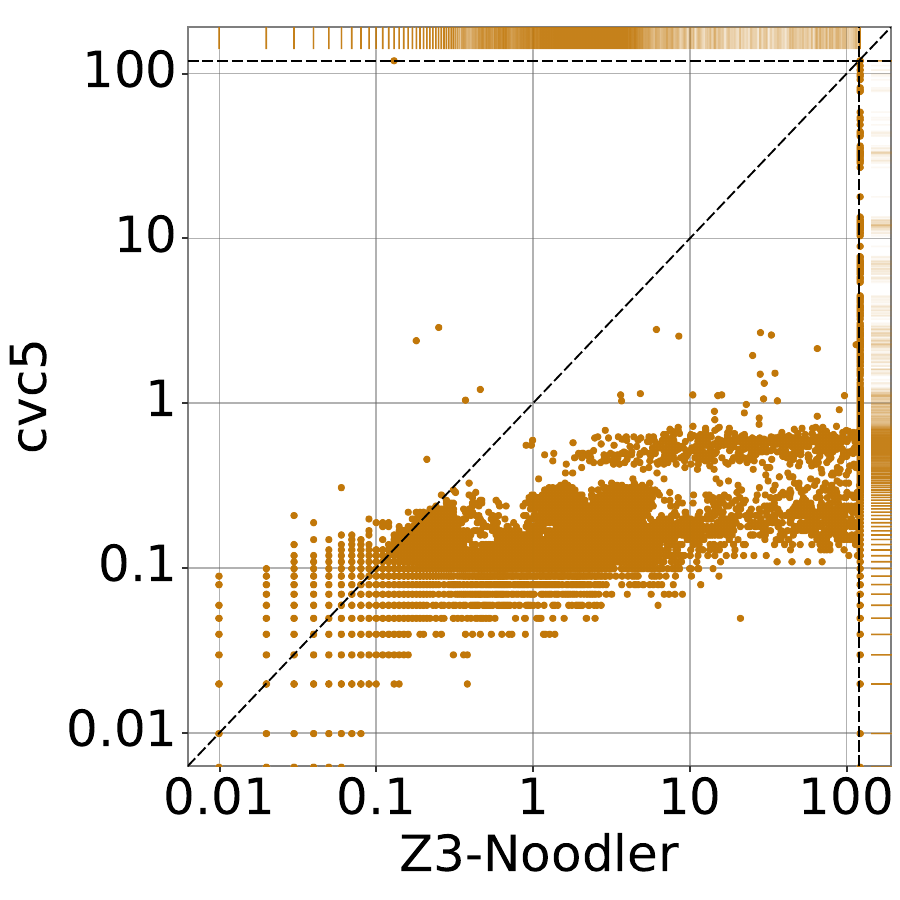}
    \caption{\ziiinoodler vs. \cvcv}
    \label{fig:cvc5}
\end{subfigure}\hfil
\begin{subfigure}{0.33\textwidth}
    \centering
    \includegraphics[width=\linewidth]{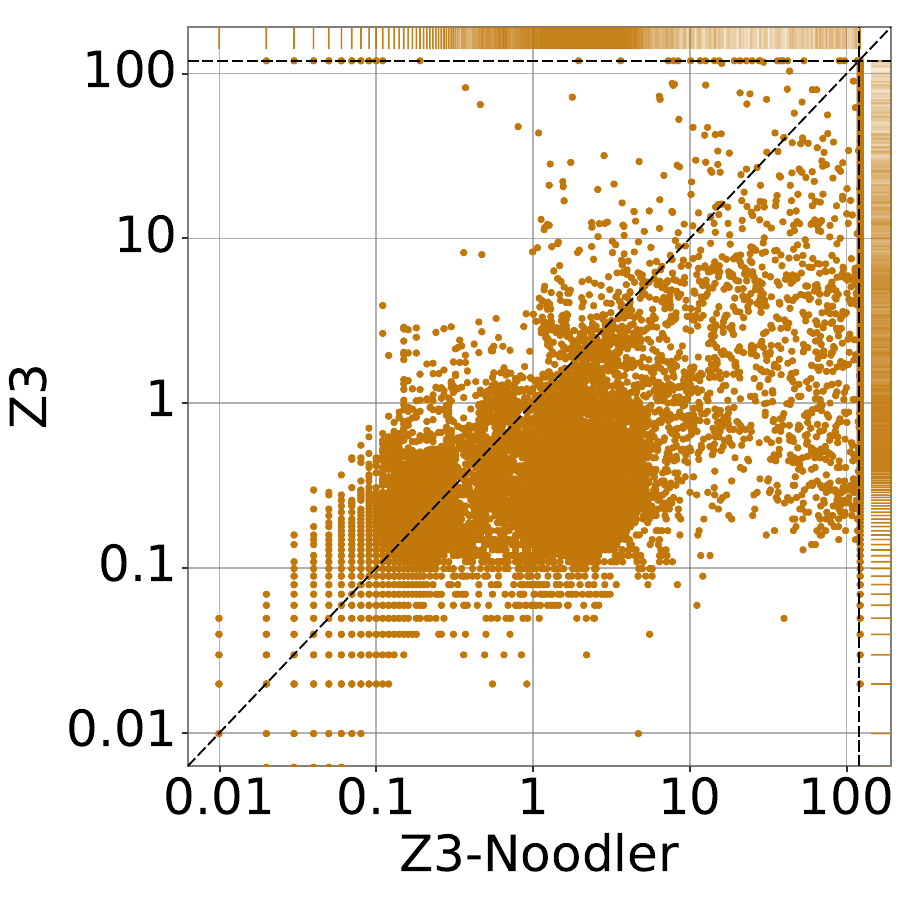}
    \caption{\ziiinoodler vs. \ziii}
    \label{fig:z3}
\end{subfigure}\hfil
\begin{subfigure}{0.33\textwidth}
    \centering
    \includegraphics[width=\linewidth]{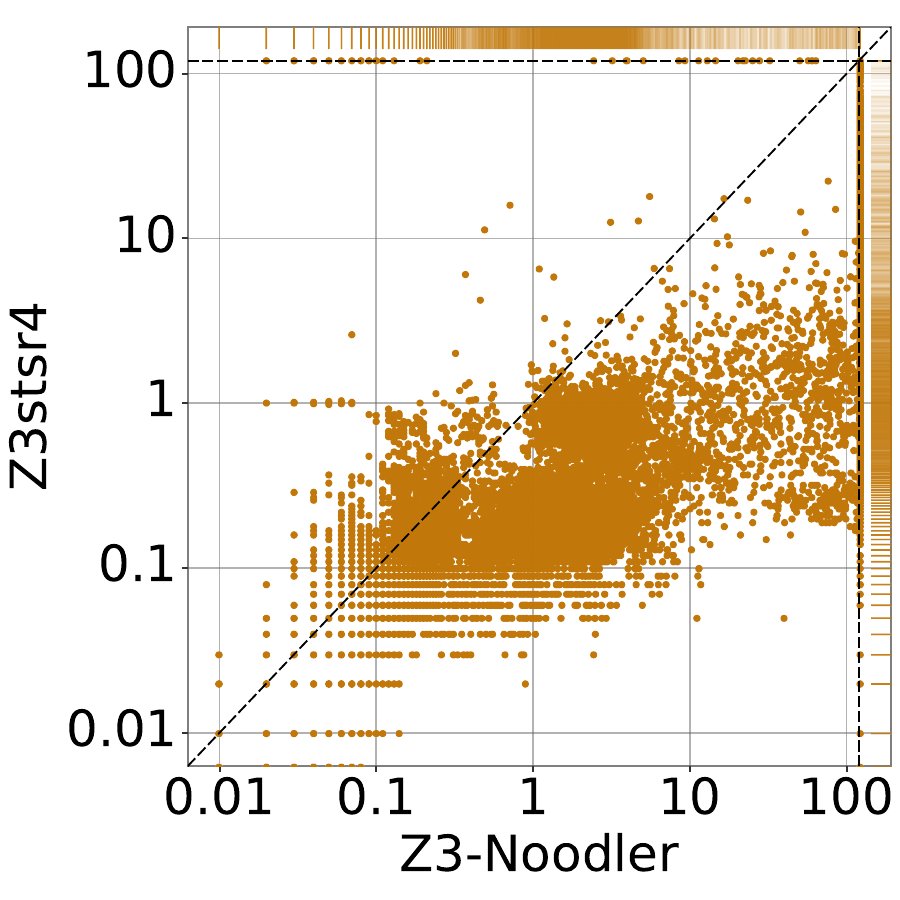}
    \caption{\ziiinoodler vs. \ziiistriv}
    \label{fig:z3str4}
\end{subfigure}

\begin{subfigure}{0.33\textwidth}
    \centering
    \includegraphics[width=\linewidth]{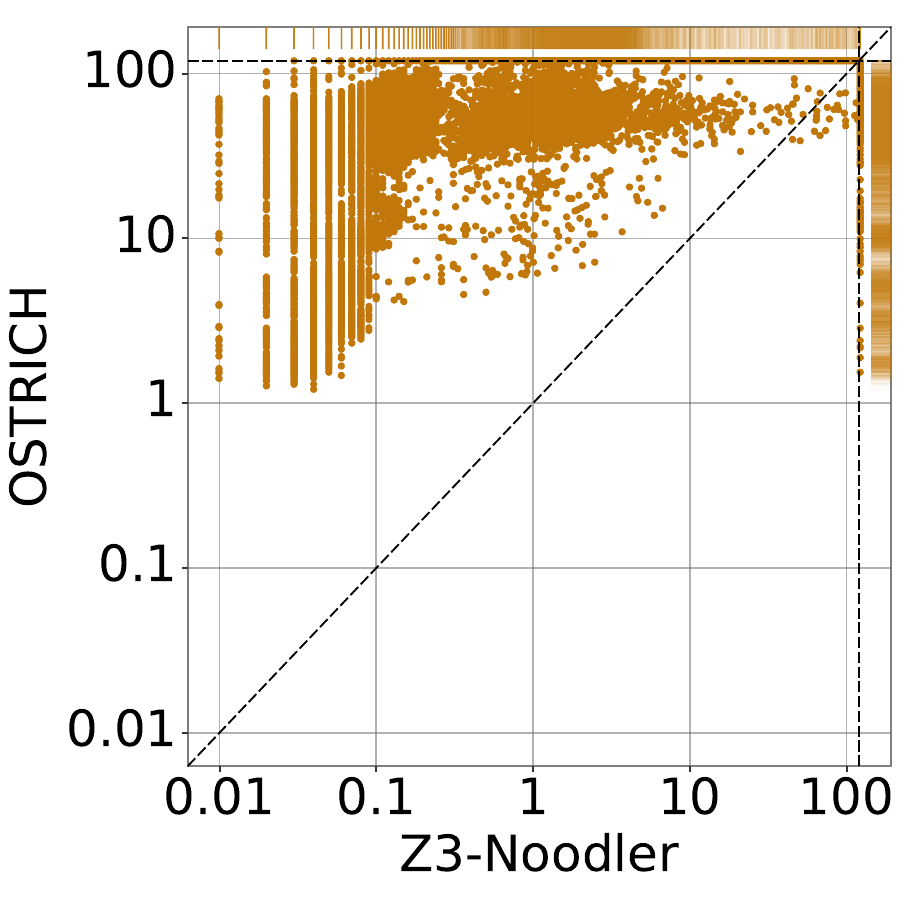}
    \caption{\ziiinoodler vs. \ostrich}
    \label{fig:ostrich}
\end{subfigure}\hfil
\begin{subfigure}{0.33\textwidth}
    \centering
    \includegraphics[width=\linewidth]{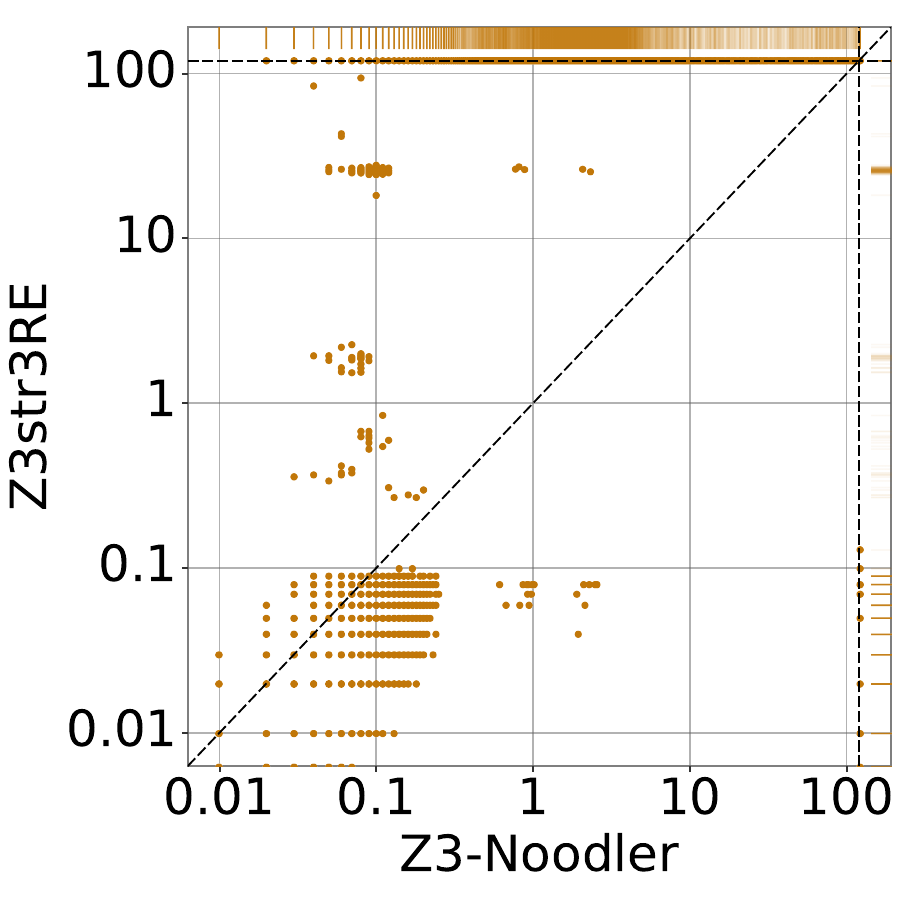}
    \caption{\ziiinoodler vs. \ziiistriiire}
    \label{fig:z3str3RE}
\end{subfigure}\hfil
\begin{subfigure}{0.33\textwidth}
    \centering
    \includegraphics[width=\linewidth]{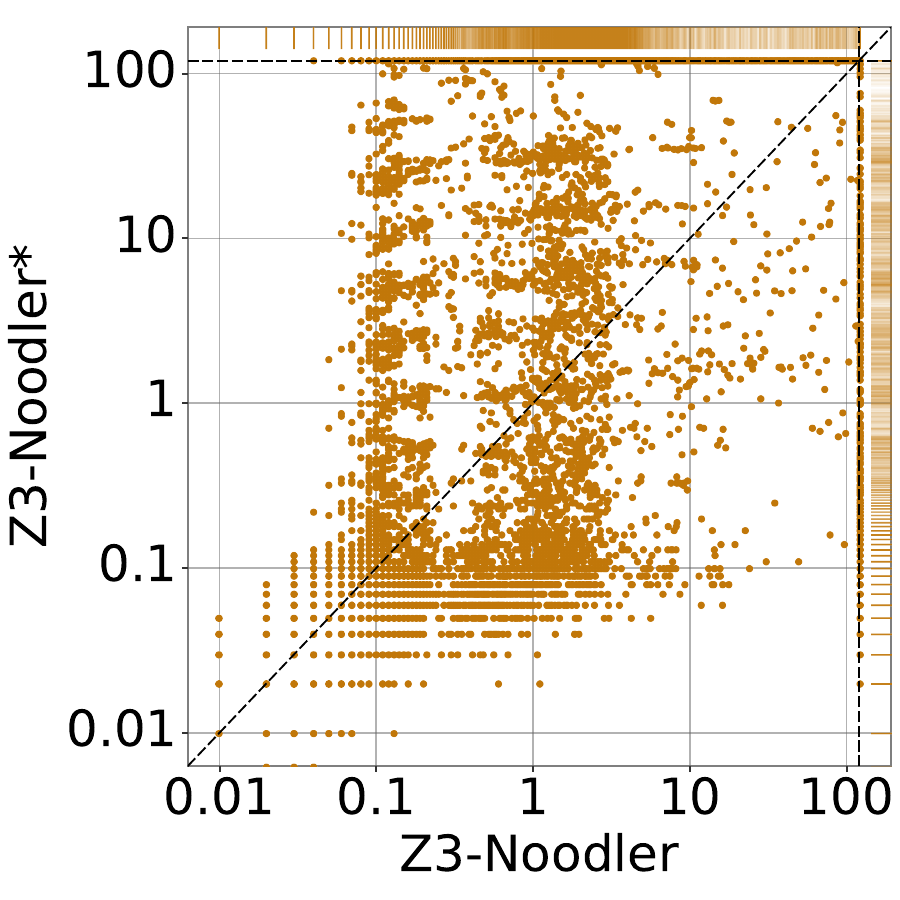}
    \caption{\ziiinoodler vs. \ziiinoodleroopsla}
    \label{fig:oldnoodler}
\end{subfigure}
\caption{
    Comparison of \ziiinoodler with other tools on \pyex benchmark.
    Times are in seconds, axes are logarithmic.
    Dashed lines represent timeouts (120\,s).
    Colors distinguish categories:
    \RGBcircle{120,10,195}\,\regexbench,
    \RGBcircle{194,119,9}\,\eqbench, and
    \RGBcircle{36,194,120}\,\predbench.
}
\label{fig:scatter-pyex}
\end{figure}

% \begin{table}[t]
% %\vspace{-1mm}
% \caption{
%     Average run times (in seconds) of solved instances and standard deviations for \pyex.
% }
% \label{tab:results-pyex}
% % \vspace{-3mm}
% % \hspace*{-4mm}
% % \resizebox{1.04\textwidth}{!}{%
% \input{figs/table_pyex.tex}%
% % \vspace{-5mm}
% % }
% \end{table}

\begin{table}
\caption{Detailed results for run times (in seconds) of the tools and the portfolio.}
\label{tab:all}
\hspace*{-4mm}
\resizebox{1.04\textwidth}{!}{%

\newcolumntype{g}{>{\columncolor{gray!30}}r}
\newcolumntype{f}{>{\columncolor{gray!30}}l}
\newcolumntype{h}{>{\columncolor{gray!30}}c}
\newcommand{\ping}[0]{\bf}
\begin{tabular}{lgggggrrrrrgggggrrrrr}
\toprule
& \multicolumn{5}{h}{\regexbench} & \multicolumn{5}{c}{\eqbench} & \multicolumn{5}{h}{\predbench} & \multicolumn{5}{c}{\pyex}\\
& \multicolumn{1}{h}{Unsolved} & \multicolumn{1}{h}{Total time} & \multicolumn{1}{h}{Avg} & \multicolumn{1}{h}{Med} & \multicolumn{1}{h}{Std} & \multicolumn{1}{c}{Unsolved} & \multicolumn{1}{c}{Total time} & \multicolumn{1}{c}{Avg} & \multicolumn{1}{c}{Med} & \multicolumn{1}{c}{Std} & \multicolumn{1}{h}{Unsolved} & \multicolumn{1}{h}{Total time} & \multicolumn{1}{h}{Avg} & \multicolumn{1}{h}{Med} & \multicolumn{1}{h}{Std} & \multicolumn{1}{c}{Unsolved} & \multicolumn{1}{c}{Total time} & \multicolumn{1}{c}{Avg} & \multicolumn{1}{c}{Med} & \multicolumn{1}{c}{Std}\\
\midrule
\rowcolor{GreenYellow}
\ziiinoodler & 62 & 2971.92 & 0.11 & 0.03 & 1.35 & 511 & 2603.78 & 0.11 & 0.02 & 2.13 & 63 & 1762.30 & 0.11 & 0.02 & 2.16 & 4424 & 65974.61 & 3.40 & 0.24 & 12.58 \\
\cvcv & 1149 & 30620.78 & 1.17 & 0.02 & 8.51 & 441 & 2840.58 & 0.11 & 0.01 & 2.15 & 4 & 505.40 & 0.03 & 0.02 & 0.15 & 34 & 7665.50 & 0.32 & 0.13 & 2.59 \\
\ziii & 571 & 51577.89 & 1.92 & 0.04 & 9.71 & 832 & 4509.83 & 0.18 & 0.02 & 2.83 & 36 & 633.79 & 0.04 & 0.03 & 0.42 & 1071 & 63069.90 & 2.77 & 0.24 & 10.54 \\
\ziiistriv & 91 & 9470.04 & 0.35 & 0.02 & 2.00 & 730 & 6002.32 & 0.25 & 0.01 & 3.40 & 46 & 291.45 & 0.02 & 0.01 & 0.31 & 570 & 33858.89 & 1.45 & 0.18 & 6.43 \\
\ostrich & 299 & 116317.79 & 4.29 & 2.30 & 8.67 & 931 & 104014.91 & 4.28 & 2.20 & 9.28 & 165 & 201112.88 & 12.71 & 6.73 & 15.08 & 12290 & 635163.13 & 54.97 & 45.37 & 78.76 \\
\ziiistriiire & 187 & 8349.97 & 0.31 & 0.01 & 3.28 & 937 & 3169.81 & 0.13 & 0.01 & 2.72 & 423 & 167.68 & 0.01 & 0.01 & 0.08 & 17764 & 4037.72 & 0.66 & 0.04 & 4.21 \\
\ziiinoodleroopsla & 1095 & 7004.06 & 0.27 & 0.04 & 2.86 & 1346 & 2826.53 & 0.12 & 0.02 & 2.93 & 344 & 1466.71 & 0.09 & 0.03 & 1.69 & 13356 & 31277.58 & 2.98 & 0.07 & 10.41 \\
\midrule
\vbsnoodler & 1 & 427.01 & 0.02 & 0.01 & 0.03 & 19 & 1303.56 & 0.05 & 0.00 & 1.62 & 2 & 162.90 & 0.01 & 0.01 & 0.14 & 24 & 7292.52 & 0.31 & 0.12 & 2.54 \\
\vbsnoodler - \ziiinoodler & 1 & 2913.72 & 0.11 & 0.01 & 0.60 & 131 & 6829.67 & 0.27 & 0.00 & 3.97 & 2 & 166.86 & 0.01 & 0.01 & 0.14 & 24 & 7297.62 & 0.31 & 0.12 & 2.54 \\
\vbsnoodler - \cvcv & 1 & 548.62 & 0.02 & 0.01 & 0.07 & 145 & 1400.79 & 0.06 & 0.00 & 1.79 & 17 & 480.98 & 0.03 & 0.01 & 1.13 & 359 & 33867.84 & 1.44 & 0.17 & 6.58 \\
\vbsnoodler - \ziii & 1 & 429.84 & 0.02 & 0.01 & 0.03 & 29 & 1579.40 & 0.06 & 0.00 & 1.87 & 2 & 164.41 & 0.01 & 0.01 & 0.14 & 27 & 7114.44 & 0.30 & 0.12 & 2.41 \\
\vbsnoodler - \ziiistriv & 1 & 473.03 & 0.02 & 0.01 & 0.04 & 19 & 1416.13 & 0.06 & 0.01 & 1.68 & 2 & 207.57 & 0.01 & 0.01 & 0.14 & 28 & 7638.42 & 0.32 & 0.13 & 2.65 \\
\vbsnoodler - \ostrich & 1 & 427.01 & 0.02 & 0.01 & 0.03 & 21 & 1269.91 & 0.05 & 0.00 & 1.61 & 2 & 162.90 & 0.01 & 0.01 & 0.14 & 24 & 7292.52 & 0.31 & 0.12 & 2.54 \\
\vbsnoodler - \ziiistriiire & 1 & 509.62 & 0.02 & 0.01 & 0.03 & 20 & 1306.84 & 0.05 & 0.01 & 1.50 & 2 & 215.62 & 0.01 & 0.01 & 0.14 & 24 & 7332.53 & 0.31 & 0.12 & 2.54 \\
\cvcv + \ziii + \ziiinoodler & 1 & 608.48 & 0.02 & 0.01 & 0.05 & 22 & 1470.72 & 0.06 & 0.01 & 1.58 & 2 & 342.24 & 0.02 & 0.02 & 0.14 & 28 & 7715.58 & 0.32 & 0.13 & 2.65 \\
\cvcv + \ziii & 278 & 27916.42 & 1.03 & 0.02 & 7.69 & 303 & 2805.47 & 0.11 & 0.01 & 2.25 & 2 & 396.73 & 0.02 & 0.02 & 0.15 & 28 & 7728.58 & 0.32 & 0.13 & 2.65 \\
\bottomrule
\end{tabular}

}
\end{table}

\end{document}